\documentclass[%
reprint,
superscriptaddress,
floatfix,
amsmath,amssymb,
aps,
prx,
longbibliography
]{revtex4-2}

\usepackage{graphicx}
\usepackage{dcolumn}
\usepackage{bm}
\usepackage{placeins}
\usepackage{float}
\usepackage{hyperref}
\usepackage{placeins}
\usepackage{times}
\usepackage{physics}
\usepackage{mathtools}
\usepackage{xcolor}
\usepackage{float}
\usepackage{hyperref}
\hypersetup{
    colorlinks,
    citecolor=blue,
    filecolor=black,
    linkcolor=blue,
    urlcolor=black
}
\usepackage[title,toc]{appendix}

\begin{document} 

\title{A dual-species Rydberg array}

\author
{\normalsize{Shraddha Anand,$^{1,\dag}$ 
Conor E. Bradley,$^{1,\dag}$
Ryan White,$^{2}$ 
Vikram Ramesh,$^{2}$ 
Kevin Singh,$^{1}$ 
and Hannes Bernien$^{1\ast}$}\\
\small{$^{1}$Pritzker School of Molecular Engineering, University of Chicago,  Chicago, IL 60637, USA}\\
\small{$^{2}$Department of Physics, University of Chicago, Chicago, IL 60637, USA}\\
\small{$^\dag$These authors contributed equally to this work.} \\
\small{$^\ast$To whom correspondence should be addressed; E-mail:~bernien@uchicago.edu.}
}

\date{\today}

\begin{abstract}
Rydberg atom arrays have emerged as a leading platform for quantum information science. Reaching system sizes of hundreds of long-lived qubits, these arrays are used for highly coherent analog quantum simulation \cite{scholl2021quantum,ebadi2021quantum}, as well as digital quantum computation \cite{graham2022multi,bluvstein2022quantum}. Advanced quantum protocols such as quantum error correction, however, require midcircuit qubit operations, including the replenishment, reset, and readout of a subset of qubits. A compelling strategy to achieve these capabilities is a dual-species architecture in which a second atomic species can be controlled without crosstalk~\cite{singh2023}, and entangled with the first via Rydberg interactions. Here, we realize a dual-species Rydberg array consisting of rubidium (Rb) and cesium (Cs) atoms, and explore new regimes of interactions and dynamics not accessible in single-species architectures. We achieve enhanced interspecies interactions by electrically tuning the Rydberg states close to a F{\"o}rster resonance \cite{beterov2015rydberg}. In this regime, we demonstrate interspecies Rydberg blockade and implement quantum state transfer from one species to another. We then generate a Bell state between Rb and Cs hyperfine qubits via an interspecies controlled-phase gate. Finally, we combine interspecies entanglement with native midcircuit readout to achieve quantum non-demolition measurement of a Rb qubit using an auxiliary Cs qubit. The techniques demonstrated here pave the way toward scalable measurement-based protocols and real-time feedback control in large-scale quantum systems.

\end{abstract}

\maketitle

\section*{Introduction}
Neutral atoms trapped in arrays of optical tweezers have recently been established as a frontrunner for both analog quantum simulation and digital quantum computation \cite{browaeys2020many,kaufman2021quantum}. The scalability of optical tweezer technologies has enabled systems of hundreds to thousands of atomic qubits \cite{ebadi2021quantum,scholl2021quantum,huft2022simple,pause2023supercharged,tao2023high}, while the development of efficient control schemes based on off-the-shelf optical components has led to the implementation of increasingly complex Hamiltonians and circuits \cite{graham2022multi, chen2023continuous, shaw2024multi,lis2023mid,bluvstein2023logical}. Together with innovative approaches to error correction based on the multi-level nature of atomic qubits \cite{wu2022erasure,cong2022hardware,ma2023high,scholl2023erasure}, these features have equipped neutral atom processors for exploring intermediate-scale quantum information science \cite{shaw2023benchmarking} and logical circuit operation \cite{bluvstein2023logical}.

The further development of this platform, however, requires solutions to outstanding challenges, which include midcircuit readout in large arrays \cite{deist2022mid,singh2023,graham2023mid,ma2023high,lis2023mid,norcia2023mid,scholl2023erasure,bluvstein2023logical} and the continuous replenishment of atoms lost during system operation \cite{singh2023,norcia2023mid,pause2023reservoir}. These are essential for operating deep circuits and for auxiliary-qubit-assisted protocols, including quantum error correction~\cite{terhal2015quantum}. A defining requirement for these challenges is that they must be performed with negligible crosstalk: `auxiliary' qubits entangled with `data' qubits must be addressed without decohering the latter.

A promising strategy to achieve this requirement is to realize a dual-species Rydberg array \cite{beterov2015rydberg,zeng2017entangling,singh2022,sheng2022defect,singh2023}, such that the unique addressing frequencies for each species result in independent control, crosstalk-free measurement \cite{singh2023}, and straightforward methods for generating Rydberg-based entanglement between them. The intrinsic addressability of qubits allows for universal control schemes \cite{cesa2023universal}, efficient state preparation \cite{muller2009mesoscopic}, and the study of novel Hamiltonians \cite{homeier2023realistic,chepiga2023tunable}, while maintaining a minimal control architecture of two separate global drives. While prior demonstrations have shown Rydberg interactions between two atoms of different isotopes \cite{zeng2017entangling}, an array of dual-species Rydberg atoms has yet to be demonstrated and would additionally benefit from using distinct atomic elements to dramatically suppress crosstalk.  

In this article, we realize a dual-species Rydberg array composed of rubidium (Rb) and cesium (Cs) atoms. We show that, by choice of the Rydberg states and control of the electric field, we can access both the van der Waals (vdW) and resonant dipole-dipole interaction regimes, the latter of which arises due to the presence of a hitherto unobserved interspecies F{\"o}rster resonance \cite{beterov2015rydberg}.
We demonstrate Rydberg blockade and show that independent control of the two species enables access to unexplored dynamical regimes in globally driven Rydberg arrays. The blockade facilitates parallel controlled-phase (CZ) gates on pairs of hyperfine qubits, allowing us to generate interspecies Bell states. Finally, we demonstrate that these entangling operations are compatible with midcircuit readout, implementing auxiliary-based quantum non-demolition (QND) measurement. 

Our results showcase the richness of interaction regimes and dynamics that can be accessed in a dual-species tweezer array architecture. Specifically, the interspecies F\"orster resonance provides anisotropic interactions for engineering many-body phases of matter~\cite{de2019observation, kim2023realization}, species-dependent interaction asymmetry for low-crosstalk and native multi-qubit gates \cite{beterov2015rydberg,mcdonnell2022demonstration,m2023parallel}, and beyond-nearest-neighbour connectivity \cite{baker2021exploiting}. 
Furthermore, the crosstalk-free and scalable control techniques offer exciting opportunities for new forms of quantum information processing~\cite{cesa2023universal}, as well as efficient methods for creating long-range entangled states via measurement and feed-forward~\cite{lu2022measurement,iqbal2023topological}.

\begin{figure*}
\includegraphics[width=\textwidth]{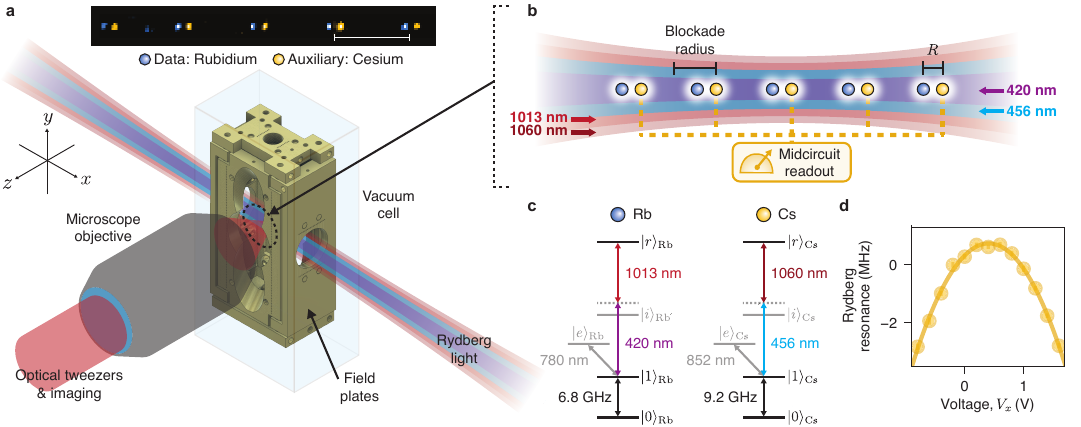}
\caption{\textbf{Rydberg interactions in a dual-species atom array.} \textbf{a}, Example fluorescence image of the Rb-Cs atom array. Interspecies pairs (Rb-Cs) are separated by $\sim$40~µm (indicated by scalebar) to suppress crosstalk between adjacent couplets. Analogously, for intraspecies experiments (Rb-Rb, Cs-Cs), pairs are separated by $\sim$25~µm. Experiments are performed inside an in-vacuum segmented Faraday cage that shields the atoms from stray electric fields and enables electric field control via application of voltages to the electrically isolated field plates. \textbf{b}, The four Rydberg laser beams are focused onto the atom array to enable high Rydberg Rabi frequencies in a counter-propagating configuration that minimizes Doppler shifts. In the Rydberg state, Rb-Cs pairs experience dipolar interactions that are leveraged for interspecies entanglement. The spectral distinguishability between the atomic species enables crosstalk-free midcircuit readout of the Cs qubits. \textbf{c}, Atoms are excited to their respective Rydberg states via species-selective two-photon transitions. The qubits are encoded either in the ground-Rydberg $\ket{1} \xleftrightarrow{} \ket{r}$, or the hyperfine $\ket{0} \xleftrightarrow{} \ket{1}$ manifold, with $\ket{0}_{\mathrm{Rb}}\,\textrm{=}\,\ket{F\,\textrm{=}\,1,m_{F}\,\textrm{=}\,0}$, $\ket{1}_{\mathrm{Rb}}\,\textrm{=}\,\ket{2,0}$, $\ket{0}_{\mathrm{Cs}}\,\textrm{=}\,\ket{3,0}$, $\ket{1}_{\mathrm{Cs}}\,\textrm{=}\,\ket{4,0}$. Qubit readout is performed via fluorescence on the $\ket{1} \xleftrightarrow{} \ket{e}$ transitions. \textbf{d}, The electric field at the position of the atoms is eliminated by applying voltages between pairs of field plates along each axis. Data shows an example Stark shift measurement of the $\ket{1}_{\mathrm{Cs}} \xleftrightarrow{} \ket{r}_{\mathrm{Cs}}$ resonance arising from a voltage $V_{x}$ applied along the x-direction.}
\label{fig:schematic}
\end{figure*}

\subsection*{Dual-species Rydberg tweezer array}

In our setup, Rb and Cs atoms are cooled, trapped, and imaged independently in optical tweezer arrays formed at the center of an ultra-high vacuum glass cell \cite{singh2022,singh2023}. The glass cell houses a segmented Faraday cage that provides electric field control of the environment (Fig.~\ref{fig:schematic}a, methods), and apertures in the metal plates of the cage provide optical access. In this work, we focus on interactions between pairs of atoms trapped in a tweezer array.

\begin{figure*}
\includegraphics[scale=1]{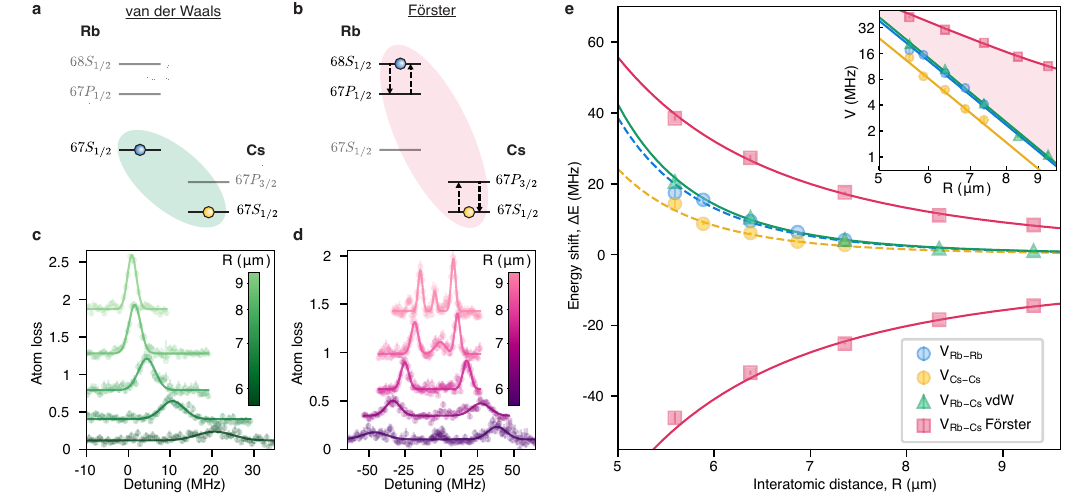}
\caption{\textbf{F{\"o}rster vs. van der Waals Rydberg interactions.} \textbf{a}, Rb-Cs pairs excited to $\ket{67S_{1/2}}_{\mathrm{Rb}}\ket{67S_{1/2}}_{\mathrm{Cs}}$ interact via second-order vdW interactions (green). \textbf{b}, Rb-Cs pairs excited to $\ket{68S_{1/2}}_{\mathrm{Rb}}\ket{67S_{1/2}}_{\mathrm{Cs}}$ undergo resonantly enhanced F{\"o}rster interactions (pink) due to near-degeneracy with the neighboring pair state $\ket{67P_{1/2}}_{\mathrm{Rb}}\ket{67P_{3/2}}_{\mathrm{Cs}}$. \textbf{c}, The vdW interaction strength is extracted from the shift in the Rb Rydberg resonance after exciting the Cs atom to the Rydberg state. The strength increases with decreasing interatomic separation. The vertical axis is offset for clarity, and the data is fit to Gaussian profiles to extract the centers of the features. \textbf{d}, The F{\"o}rster interaction strength is extracted similarly but reveals additional features; the two main peaks correspond to the eigenstates $\ket{\pm}_{\text{pair}}$ (see main text). The smaller peaks at zero detuning correspond to the erroneous cases where the Cs atom was not excited to the Rydberg state. The center peak at larger spacings stems from additional resonant pair states, which can be suppressed by tuning the electric field (methods). \textbf{e}, Measured energy shifts are plotted against interatomic spacing for the Rb-Rb, Cs-Cs, Rb-Cs vdW, and the Rb-Cs F{\"o}rster interactions. The theoretically predicted (dashed) curves for the homogeneous pairs are used to calibrate the x-axis. By fitting the heterogeneous pair data to the appropriate functional forms (solid lines, methods), we find $C_{3}\,\textrm{=}\,$16.4(3)~GHz~µm$^3$, and $C_{6}\,\textrm{=}\,$662(21)~GHz~µm$^6$, both of which are compatible with theoretically predicted values. The inset shows the interaction strengths plotted on a log-log scale, defining $V\,\textrm{=}\,\Delta E$ for the vdW interactions, and $V\,\textrm{=}\,(\Delta E_\text{u}-\Delta E_\text{l})/2$ for the F\"orster interaction, where $\Delta E_\text{u(l)}$ denotes the energy of the upper (lower) branch.} 
\label{fig:spectroscopy}
\end{figure*}

Atoms are initialized in the hyperfine clock states $\ket{1}_{\mathrm{Rb}}$ and $\ket{1}_{\mathrm{Cs}}$. From here, each species can be independently excited to high-lying Rydberg states via species-selective two-photon transitions (Fig.~\ref{fig:schematic}b,c). We excite to the $m_{j}\,\textrm{=}\,1/2$ Rydberg states using frequency selectivity and polarization control (methods). The four Rydberg lasers are simultaneously locked to a single ultra-low expansion cavity to achieve narrow linewidths and suppress phase noise (methods). Since the laser frequencies that address these transitions are far-detuned from one another, the principal quantum number for the Rydberg states $\ket{r}_{\mathrm{Rb}}$ and $\ket{r}_{\mathrm{Cs}}$ can be chosen independently for the two species. This choice determines whether the atoms interact via dipole-dipole or van der Waals interactions (see next section). Qubits are encoded either in the ground-Rydberg (gr) manifold $(\ket{1} \xleftrightarrow{} \ket{r})$, or in the hyperfine (hf) manifold $(\ket{0} \xleftrightarrow{} \ket{1}$). In the latter case, selective Rydberg excitation is still performed on the $\ket{1} \xleftrightarrow{} \ket{r}$ transition, but single-qubit operations between the qubit states $\ket{0}$ and $\ket{1}$ are implemented via microwave driving \cite{singh2023}. During Rydberg operations, the tweezer light is turned off as Rydberg atoms experience an anti-trapping potential from the tweezers~\cite{de2018analysis}. In the case of gr-encoding, the ramping up of tweezers is thus used to eject atoms in the Rydberg state: the remaining ground-state qubits can be detected via atomic fluorescence. Similarly, for the hf encoding, resonant light pulses (`pushout') are used to selectively remove atoms in $\ket{1}$. Finally, midcircuit hyperfine readout can be performed on a subset of the qubits while dynamically decoupling the others (methods).

\subsection*{Rydberg spectroscopy}

We begin by characterizing the Rydberg interaction strength between homogeneous and heterogeneous pairs of atoms. For the case of Rb-Rb $(\ket{r}_{\mathrm{Rb}}\,\textrm{=}\,\ket{68S_{1/2}}\,\textrm{:=}\,\ket{68})$, Cs-Cs $(\ket{r}_{\mathrm{Cs}}\,\textrm{=}\,\ket{67S_{1/2}}\,\textrm{:=}\, \ket{67})$, and Rb-Cs pairs with the same principal quantum number $(\ket{67, 67}_{\mathrm{Rb, Cs}})$, we expect van der Waals-type interactions that scale as $1/R^{6}$, where $R$ is the interatomic spacing. Although the Rydberg interaction is fundamentally dipolar, this second-order scaling occurs due to significant non-degeneracy of Rydberg pair-states~\cite{browaeys2016experimental} and is the typical regime used in atom arrays (Fig.~\ref{fig:spectroscopy}a).  For the interspecies case of $\ket{68,67}_{\mathrm{Rb, Cs}}$, however, there is a predicted coincidental near-degeneracy with another pair-state, $\ket{67P_{1/2},67P_{3/2}}_{\mathrm{Rb, Cs}}$, called a F{\"o}rster resonance (Fig.~\ref{fig:spectroscopy}b), which would result in the re-emergence of resonant dipole-dipole interactions ($1/R^{3}$)~\cite{beterov2015rydberg}. Since the resonance is sensitive to electric fields (methods), we use the Faraday cage to nullify stray fields at the position of the atoms. This is achieved by maximizing the Rydberg transition energy (Fig.~\ref{fig:schematic}d).

For each combination of atomic states, we perform spectroscopy for multiple interatomic spacings ranging from $5.6$~µm to $9.3$~µm. In Fig.~\ref{fig:spectroscopy}c,d, we show the measured spectra when exciting on the $\ket{67,67}_{\mathrm{Rb, Cs}}$ and $\ket{68,67}_{\mathrm{Rb, Cs}}$ transitions, respectively. We observe qualitatively distinct behavior. In particular, a clear splitting of the spectrum is observed for the $\ket{68,67}_{\mathrm{Rb, Cs}}$ case. The two branches in the spectrum, corresponding approximately to the eigenstates
\begin{equation*}
\ket{\pm}_{\text{pair}} = \ket{68S_{1/2}}_{\mathrm{Rb}}\ket{67S_{1/2}}_{\mathrm{Cs}} \pm \ket{67P_{1/2}}_{\mathrm{Rb}}\ket{67P_{3/2}}_{\mathrm{Cs}},
\end{equation*}
confirm the presence of the near-degenerate F{\"o}rster resonance \cite{comparat2010dipole,beterov2015rydberg,ravets2014coherent,chew2022ultrafast}. In both cases, the interaction strength increases with proximity. A broadening of the resonances arises due to positional fluctuations from the finite temperature of the atoms~\cite{ravets2014coherent,bernien2017probing}.

\begin{figure}
\includegraphics[scale=.95]{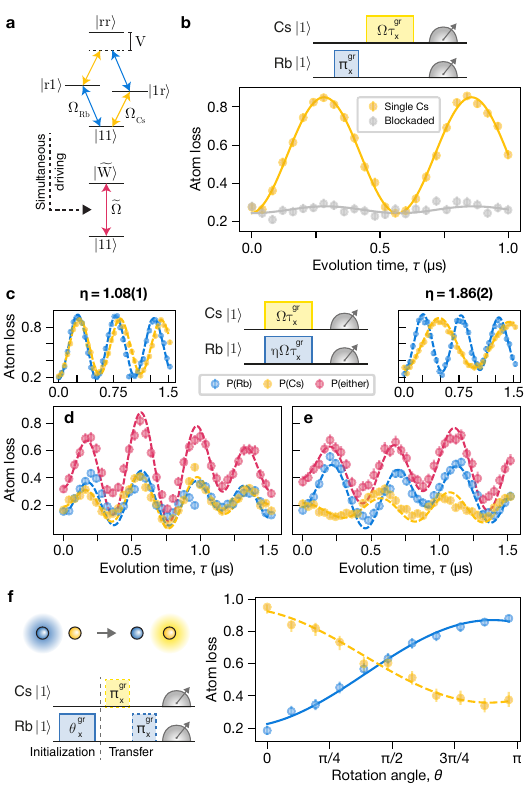}
\caption{\textbf{Dual-species Rydberg blockade and dynamics.} \textbf{a}, Simplified Rb-Cs ground-Rydberg (`gr') manifold level diagram depicting that the doubly-excited state shifts out of resonance due to the Rydberg blockade. Under simultaneous driving, the ground state couples to a singly-excited $W$-like state $|\widetilde{W}\rangle$, with enhanced Rabi frequency $\widetilde{\Omega}$ (methods). \textbf{b}, Rb atoms are prepared in $\ket{r}$, followed by Rabi driving of the Cs atoms. Conditioning on single vs. pair loading and the correct preparation of the Rb atom (methods), the fitted Cs oscillation amplitude is strongly suppressed from $A\,\textrm{=}\,0.606(18)$ to $A\,\textrm{=}\,0.036(24)$. \textbf{c}, In the absence of the other species, Rb and Cs exhibit typical Rydberg Rabi oscillations. \textbf{d}, \textbf{e}, Observed and simulated Rydberg dynamics under simultaneous driving. \textbf{d}, The Rabi frequencies are balanced such that $\eta \,\textrm{=}\, \Omega_{\textrm{Rb}}/\Omega_{\textrm{Cs}}\,\textrm{=}\,1.92(1)/1.78(1)\,\textrm{=}\,1.08(1)$. \textbf{e}, 
The Rabi frequencies are unbalanced such that $\eta\,\textrm{=}\,1.93(1)/1.04(1)\,\textrm{=}\,1.86(2)$. When paired, enhanced oscillations occur at 2.53(2)~MHz (panel d) and 2.17(2)~MHz (panel e), as expected from  $\widetilde{\Omega}$. The relative excitation probability is theoretically given by $\Omega_{i}^{2}/\widetilde{\Omega}^{2}, i \in \{\textrm{Rb},\textrm{Cs}\}$. These observations are affected by SPAM errors but well-described by numerical simulations taking these into account (dashed lines, panels c-e). \textbf{f}, For the interspecies quantum state transfer, the Rb atom is initialized in an arbitrary state via $R_x(\theta)$ (solid blue line). After initialization, a $\pi$-pulse on Cs generates $\langle ZZ \rangle$ correlations, and a subsequent $\pi$-pulse on Rb disentangles the atoms and completes the transfer (dashed yellow line).} 
\label{fig:groundrydberg}
\end{figure}

When plotted against interatomic spacing in Fig.~\ref{fig:spectroscopy}e, we observe that, indeed, the Rb-Cs F{\"o}rster interaction results in a slower fall-off than the Rb-Cs vdW interactions. By fitting the curves with the expected $C_{3}/R^3$ and $C_{6}/R^6$ functional forms (methods), we extract $C_3\,\textrm{=}\,$16.4(3)~GHz~µm$^3$, and $C_6\,\textrm{=}\,$662(21)~GHz~µm$^6$, in agreement with theoretically predicted values of $C_{3}\,\textrm{=}\,$15.66(2)~GHz~µm$^3$, and $C_{6}\,\textrm{=}\,$745(1)~GHz~µm$^6$ \cite{beterov2015rydberg,Weber2017}. The interatomic spacing for the interspecies interactions was calibrated by fitting the measured intraspecies vdW interaction curves in Fig.~\ref{fig:spectroscopy}e with theoretically calculated values of $C_6$. Due to the isotropy and robustness to electric field noise of the vdW interactions and the availability of experimentally-verified numerical tools for same-species interactions~\cite{Weber2017}, this method provides a well-calibrated ruler for the distance between atoms.  

These measurements constitute the first observation of an interspecies F{\"o}rster resonance between individually trapped atoms. The existence of such a resonance gives rise to a number of features including long-range interactions \cite{bornet2023scalable,chen2023continuous}, anisotropy \cite{ravets2015measurement,de2019observation}, and tunable inter:intra-species interaction asymmetry (methods) \cite{beterov2015rydberg}. 
Indeed, even at modest trap spacings of 9.3~µm, we observe an asymmetry factor $>$10  between the interspecies F\"orster and intraspecies van der Waals interactions (Fig.~\ref{fig:spectroscopy}e, inset).

\subsection*{Interspecies Rydberg blockade and dynamics}

Operating at the Rb-Cs F{\"o}rster pair resonance, we next employ coherent control of the ground-Rydberg qubit manifold to realize interspecies blockade. For these experiments, pairs of atoms are placed at a distance of 5.6~µm. This results in an effective interspecies interaction strength of $\sim$24~MHz (methods), for which we expect strong blockade dynamics, i.e. the doubly-excited state $\ket{rr}$ is shifted out of resonance (Fig.~\ref{fig:groundrydberg}a).

To experimentally verify this, we make use of the independent addressability of the two species. We first apply a $\pi$-pulse to the Rb qubits to excite them to $\ket{r}_{\mathrm{Rb}}$, and then attempt to drive Rabi oscillations on the Cs $\ket{1}_{\mathrm{Cs}} \xleftrightarrow{} \ket{r}_{\mathrm{Cs}}$ transition (Fig.~\ref{fig:groundrydberg}b). The collected data is post-selected on loading pairs of atoms and on successful state preparation of the Rb qubit (methods). We find that the amplitude of the Cs Rabi oscillation is dramatically suppressed in the presence of a Rb atom, a clear manifestation of the Rydberg blockade. The data presented here is not corrected for the remaining state-preparation and measurement (SPAM) errors, which are the dominant limitations on the contrast of the Rabi oscillations. To minimize contributions from atoms not in $\ket{1}$, we convert these errors to loss (erasure) by state-selective pushout prior to any Rydberg driving (methods).

A second indicator of blockade physics is the observation of enhanced Rabi oscillations between the fully-occupied ground state and a collective singly-excited state \cite{gaetan2009observation,zeiher2015microscopic,labuhn2016tunable,bernien2017probing}. For $N$ blockaded atoms undergoing collective driving, the single-atom Rabi frequency is enhanced by $\sqrt{N}$, while the maximal single-atom excitation probability is given by $1/N$. With independent drives on each species, however, a richer variety of dynamics can be accessed. Considering two atoms with Rabi frequencies $\Omega_{1}$ and $\Omega_{2}$, enhanced oscillations now occur at a collective frequency $\widetilde{\Omega}\,\textrm{=}\,\sqrt{\Omega_{1}^{2}+\Omega_{2}^{2}}$, while the maximal excitation probability on each site is given by $\Omega_{i}^{2}/\widetilde{\Omega}^{2}, i \in \{1,2\}$ (methods). Here, we demonstrate these phenomena.

\begin{figure*}
\includegraphics[scale=1]{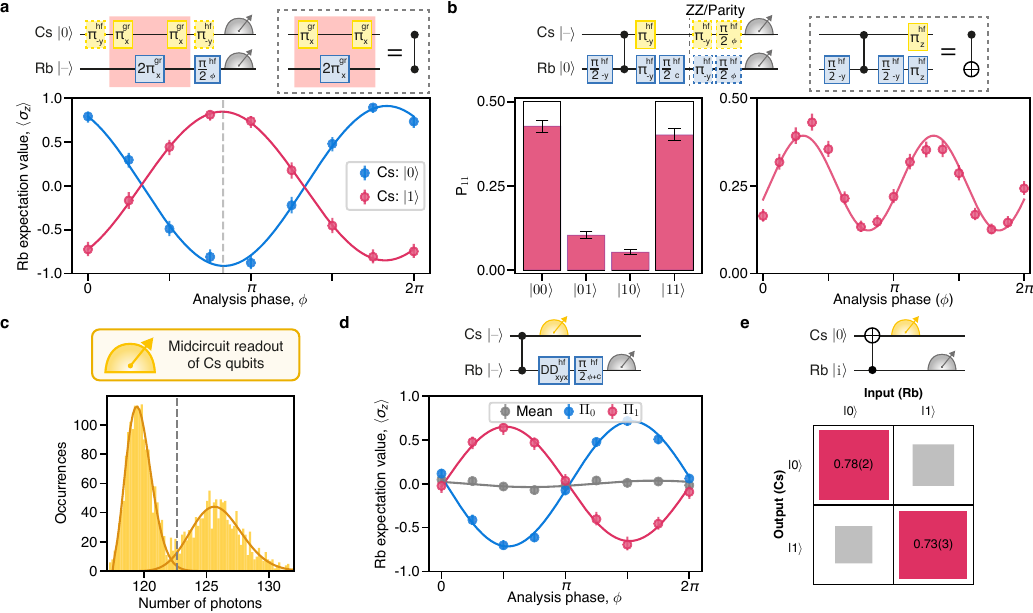}
\caption{\textbf{Entanglement of long-lived hyperfine qubits and midcircuit QND readout}. \textbf{a}, Measurement of the conditional phase accrued by Rb atoms undergoing a $\pi$-2$\pi$-$\pi$ sequence (`eye-diagram'). The Cs atoms are prepared in the eigenstates $\ket{0}$ or $\ket{1}$ (dashed boxes). The conditional phase is 1.01(1)$\pi$, and the SPAM-corrected contrast is 0.88(1). The gr pulses implement a hyperfine-CZ gate up to single-qubit phases. Dashed gray line indicates the point of maximal contrast. Spin echoes (`$\pi^{\text{hf}}_\text{-y}$') are incorporated to mitigate hyperfine dephasing. \textbf{b}, Parallel preparation of the canonical Bell state $|\Psi^{+}\rangle$ across multiple Rb-Cs pairs. By measurement of the populations and coherences (dashed boxes) we extract an average SPAM-corrected Bell state fidelity of $\mathcal{F}_{\mathrm{Bell}}\,\textrm{=}\,0.69(3)$. The combination of hf $\pi/2$-pulses and gr pulses realizes a CNOT operation up to single-qubit rotations. The phase correction `$c$' corresponds to the point of maximal contrast in panel a. \textbf{c}, Example Cs MCR histogram as used in panels d, e (7~ms fluorescence, methods); the fitted discrimination fidelity is 0.990(3). \textbf{d}, After preparing the Bell state $|\Psi^\mathrm{x}\rangle$ (see text), MCR of the Cs atoms projects the two-qubit state. Conditioning on the measurement outcomes reveals $\langle ZX \rangle_\mathrm{Cs,Rb}$ correlations, confirming that the Rb coherence survives MCR. The conditional phase is 1.02(1)$\pi$, and the SPAM-corrected contrast is 0.68(1). `DD' denotes dynamical-decoupling. \textbf{e}, Auxiliary-based QND measurement of the Rb qubit via the Cs qubit. The readout fidelity is extracted to be $\mathcal{F}_{\mathrm{QND}}\,\textrm{=}\,0.76(2)$.}
\label{fig:hyperfine}
\end{figure*}

Figure~\ref{fig:groundrydberg}c shows the single-atom $\ket{1} \xleftrightarrow{} \ket{r}$ Rabi oscillations. In contrast, in Fig.~\ref{fig:groundrydberg}d,e, we perform collective driving on pairs of Rb and Cs qubits. For the interspecies case of Rb-Cs pairs, we vary the ratio of the Rabi frequencies. When the Rabi frequencies are balanced to within 10\% ($\eta \,\textrm{=}\, \Omega_{\textrm{Rb}}/\Omega_{\textrm{Cs}} \,\textrm{=}\, 1.08(1)$, Fig.~\ref{fig:groundrydberg}d), we observe the expected $\widetilde{\Omega}$, and find that the excitation probabilities are comparable, with a minor enhancement in favor of Rb. Conversely, for a significantly imbalanced case ($\eta \,\textrm{=}\, 1.86(2)$, Fig.~\ref{fig:groundrydberg}e), it is significantly more likely that the excitation is found on the Rb atom. For Rb-Rb and Cs-Cs pairs, we recover the expected $\sqrt{2}$ enhancement in the Rabi frequency, with an accompanying reduction in the single-site excitation probability (methods). Incorporating SPAM errors, our data is well-described by numerical simulations without any free parameters (methods). We expect a significant reduction in SPAM errors with future experimental upgrades. 

Having established the interspecies Rydberg blockade, we show that this enables novel approaches to multi-qubit protocols, in particular, quantum state transfer. Here, a variable rotation $R_x(\theta)$ prepares an arbitrary Rb superposition state (Fig.~\ref{fig:groundrydberg}f, `Initialization'). Then, $\pi$-pulses are applied to the Cs and Rb qubits in turn. In the blockade regime, this pulse sequence results in the transfer of the Rb state to the Cs qubit (Fig.~\ref{fig:groundrydberg}f, `Transfer'), as indicated here by the generation of $\langle ZZ \rangle$ correlations (methods). Species-selective Rydberg excitation thus enables simple pulse schemes for quantum information processing, all within an architecture composed of global Rydberg drives. For instance, such techniques can be used for auxiliary-enhanced Rydberg state detection and for controlling the spatial flow of quantum information~\cite{cesa2023universal}.

These demonstrations highlight the unexplored dynamical regimes and information processing protocols available in collectively driven dual-species Rydberg arrays~\cite{homeier2023realistic,chepiga2023tunable, cesa2023universal}. With straightforward adjustments of the array geometry, laser detunings, or laser intensities, subsets of the atomic array can be engineered to have different dynamics from the rest  without the need for local addressing \cite{chen2023continuous, omran2019generation}.

\subsection*{Interspecies two-qubit gate and midcircuit readout}

We now extend the quantum correlations generated by the Rb-Cs blockade to the hyperfine qubit manifold and demonstrate the first interspecies logic operations in an atom array. To perform the entangling operations, we use the `$\pi$-2$\pi$-$\pi$' protocol~\cite{jaksch2000fast}, with the $\pi$-pulses performed on Cs and the intermediary 2$\pi$-pulse performed on Rb. If the first $\pi$-pulse excites the Cs qubit to the Rydberg state, the Rb qubit is blockaded and remains unaffected; otherwise, it acquires a geometric phase of $\pi$. This phenomenon allows us to engineer a controlled-phase (CZ) gate. To verify this, the Rb qubits are prepared in the superposition state $\ket{-}_{\mathrm{Rb}}$, the Cs qubits are prepared in each of the two eigenstates $\ket{0}_{\mathrm{Cs}}$ or $\ket{1}_{\mathrm{Cs}}$, and the $\pi$-2$\pi$-$\pi$ scheme is applied. Measuring the phase accrued by the Rb qubits results in an `eye diagram’ (Fig.~\ref{fig:hyperfine}a)  with a conditional phase of 1.01(1)$\pi$ as desired. Throughout these hf-manifold measurements, we build on Refs. \cite{mcdonnell2022demonstration,ma2023high} to develop a straightforward method for the correction of SPAM errors at the cost of reduced raw fidelities by converting the majority of these to loss (methods).

With the interspecies two-qubit gate we can now generate a maximally entangled Bell state. We prepare both species in the $\ket{-}$ state, perform the entangling sequence, and then apply a single-qubit $R_{\phi}(\pi/2)$ gate on the Rb qubit, with $\phi$ set to the point of maximum contrast in Fig.~\ref{fig:hyperfine}a. Asessing the Bell state via measurement of the two-qubit populations and coherences (Fig.~\ref{fig:hyperfine}b), we clearly observe the expected correlations in the populations alongside the enhanced frequency associated with parity oscillations of an entangled state. The Bell state fidelity is evaluated to be $\mathcal{F}_{\mathrm{Bell}}\,\textrm{=}\,0.69(3)$ after correction of SPAM errors (raw fidelity $\mathcal{F}^{\mathrm{raw}}_{\mathrm{Bell}}\,\textrm{=}\,0.49(2)$). We thus demonstrate Rydberg blockade-based entanglement of two different atomic species. The measured SPAM-corrected fidelity is in good agreement with our error model, which predicts a value of $\mathcal{F}^{\mathrm{sim}}_{\mathrm{Bell}}\,\textrm{=}\,0.73$ (methods). This is predominantly limited by two dephasing mechanisms: ground-Rydberg $T^{*}_{2}$ of the idling Cs atom (estimated infidelity $\epsilon^{\mathrm{gr}}_{T^{*}_{2}}\,\textrm{=}\,0.20$), and hf-manifold dephasing of both species from differential Stark-shifts induced by the blue light ($\epsilon^{\mathrm{hf}}_{\mathrm{Stark}}\,\textrm{=}\,0.04$). The latter would be suppressed by balancing the single-photon Rabi frequencies, while the former is likely limited by laser phase noise which can be addressed by technical upgrades \cite{de2018analysis,endo2018residual,li2022active,chao2023pound} or by using continuous-driving gate schemes \cite{levine2019parallel,jandura2022time}. Such technical improvements will enable higher gate fidelities, comparable to those achieved recently in both alkaline and alkaline-earth atom arrays \cite{fu2022high,evered2023high,ma2023high}.

In a final pair of measurements, we combine interspecies two-qubit gates with the midcircuit readout (MCR) capabilities of the dual-species architecture~(Fig.~\ref{fig:hyperfine}c) \cite{singh2023}. First, we demonstrate the ability to perform projective measurements on a sub-component of an entangled state. After preparing the Bell state $\ket{\Psi^\mathrm{x}}_\mathrm{Cs,Rb} \,\textrm{:=}\, (\ket{0+} + \ket{1-})/\sqrt{2}$, we perform MCR on the Cs qubit, during which the Rb qubit is decoupled. Once the MCR is complete, we measure the Rb coherence by sweeping the phase of a final $\pi/2$-pulse (Fig.~\ref{fig:hyperfine}d). Conditioning on the MCR outcomes reveals the coherent oscillations associated with the eye-diagram of Fig.~\ref{fig:hyperfine}a. Hence, the Cs MCR projects the Rb state without any additional decoherence.

Second, we implement quantum non-demolition detection of the Rb state by using Cs as an auxiliary qubit. The quantum circuit is shown in Fig.~\ref{fig:hyperfine}e: after preparing the Rb qubit in a chosen eigenstate, we insert the CZ gate between a pair of $\pi/2$-pulses on the Cs qubit, perform MCR of Cs, and then read out the Rb state. The conditional phase induced by the CZ gate correlates the Cs readout outcome with the Rb state, which is minimally perturbed in the process (measured QND-ness 0.94(2), methods). We extract a QND readout fidelity of $\mathcal{F}_{\mathrm{QND}}\,\textrm{=}\,(\mathcal{F}_{0|0}+\mathcal{F}_{1|1})/2\,\textrm{=}\, 0.76(2)$, in good agreement with our numerical model (predicted $\mathcal{F}_{\mathrm{QND}}\,\textrm{=}\,0.78$). 

The first demonstration of interspecies gates and entanglement in an optical tweezer array presented here is an essential prerequisite for advanced quantum algorithms in a dual-species quantum processor. Together with the extension of the auxiliary-qubit-based QND protocol to larger numbers of qubits, our approach will enable the exploration of quantum feedback control, error correction \cite{terhal2015quantum}, measurement-based state preparation \cite{iqbal2023topological,lu2022measurement}, and the measurement-based paradigm of quantum computing \cite{briegel2009measurement}. 

\section*{Discussion}

This work introduces the dual-species Rydberg architecture and establishes it as a powerful tool both for quantum information science and for accessing novel regimes in many-body physics. By integrating strong Rydberg interactions with the native addressability intrinsic to this approach, we have shown that a dual-species system is a natural setting in which to implement auxiliary-qubit-based protocols such as QND measurement. In combination with our recent demonstration of midcircuit feed-forward and readout in 2D arrays, and of coherent replenishment of atoms during circuit operation \cite{singh2023}, dual-species Rydberg arrays are poised to explore quantum error correction and the generation of long-range entangled states \cite{lu2022measurement,iqbal2023topological} in large systems. While improvements in operation fidelities will be required, the necessary technical upgrades are well-understood \cite{de2018analysis,evered2023high}. The further addition of fast, Raman-based single-qubit operations \cite{levine2022dispersive}, local-addressing techniques \cite{graham2022multi, chen2023continuous, shaw2024multi,lis2023mid,bluvstein2023logical}, and non-destructive read-out \cite{martinez2017fast,kwon2017parallel} will enable increasingly complex protocols.

Additionally, the independent addressability of the two species and the richness of interspecies Rydberg physics together provide means to both extend the set of implementable Hamiltonians \cite{chepiga2023tunable,homeier2023realistic}, and prepare a wide range of quantum states that are challenging to access in globally-driven single-species arrays. The experimental confirmation of inter-element Rydberg F{\"o}rster resonances offers long-range interactions \cite{defenu2023long} and the ability to tune inter:intra-species interaction asymmetry. This, in turn, gives rise to strategies for modifying qubit connectivity, exploring complex models in quantum many-body physics \cite{de2019observation, kim2023realization}, and directly implementing multi-qubit gates \cite{mcdonnell2022demonstration,m2023parallel,muller2009mesoscopic}. The F{\"o}rster interaction itself can be switched on and off on ns-timescales using the electric field control available in our setup \cite{ravets2014coherent}, which could be utilized for Floquet engineering \cite{choi2020robust,scholl2022microwave}. The dual-species neutral atom platform thus has a dual advantage: it provides access to both a wider variety of quantum phenomena and a versatile control toolbox to leverage them for quantum information science.

\section*{Acknowledgments}
We thank Sebastian Weber for helpful discussions regarding the use of the Pairinteraction software package and thank Harry Levine for critical reading of the manuscript.
We acknowledge funding from the Office of Naval Research
(N00014-23-1-2540), the Air Force Office of Scientific Research (FA9550-21-1-0209,  22-RI-EP-19), the NSF QLCI for Hybrid Quantum Architectures and Networks (NSF award 2016136). This material is based upon work supported by the U.S. Department of Energy Office of Science National Quantum Information Science Research Centers. R. W. is supported by the National Science Foundation Graduate Research Fellowship under Grant No. 2140001.

\bibliography{DualSpecies}

\onecolumngrid
\begin{appendices}

\section*{Methods}

\renewcommand{\figurename}{Extended Data Fig.}
\renewcommand{\thefigure}{\arabic{figure}}
\setcounter{figure}{0}
\renewcommand{\theHtable}{Supplement.\thetable}
\renewcommand{\theHfigure}{Supplement.\thefigure}

\subsection{Experimental platform}

As in previous work \cite{singh2023}, individual Rb and Cs atoms are loaded from a bi-chromatic 3D magneto-optical trap into species-selective optical tweezer arrays (840.6~nm for Rb, 911.3~nm for Cs) generated by independent spatial light modulators (SLMs). The tweezer arrays are focused by a high-NA objective to waists $\sim$0.9~µm, and individually homogenized for uniformity of trap depths \cite{singh2022}. The relative alignment of the two SLM arrays is optimized via measurements of site-wise Stark shifts induced by the first array on the atoms trapped in the second array. The optimization is performed by feedback on mirror positions and the tip, tilt, and blazed grating parameters associated with the SLM phase patterns. In particular, the ``defocus" Zernike polynomial allows improved co-alignment in the axial direction. 

The tweezer depths are set to $\sim$1~mK for Rb and $\sim$1.8~mK for Cs during loading, and lowered for state initialization into the $\ket{1}_{\mathrm{Rb}} \,\textrm{=}\, \ket{F\,\textrm{=}\,2,m_{F}\,\textrm{=}\,0}$ and $\ket{1}_{\mathrm{Cs}} \,\textrm{=}\, \ket{F\,\textrm{=}\,4, m_{F}\,\textrm{=}\,0}$ hyperfine clock states via $\pi$-polarized optical pumping. Single-qubit manipulations in the hf-manifold are performed using a home-built microwave horn \cite{singh2023}. Following pumping, for the interspecies measurements, we transfer the pairs of atoms to a third tweezer array generated by crossed acousto-optic deflectors (AODs) at 911.3~nm. While this co-trapping approach reduces sensitivity to any sub-micron-scale misalignment between the separate trapping arrays, it is not required for qubit operations in the strongly blockaded regime, which is robust to small positional fluctuations of the atoms. For same-species experiments, we load directly into the tweezer array generated by AODs.

\begin{figure*}[h]
\includegraphics[width=0.85\textwidth]{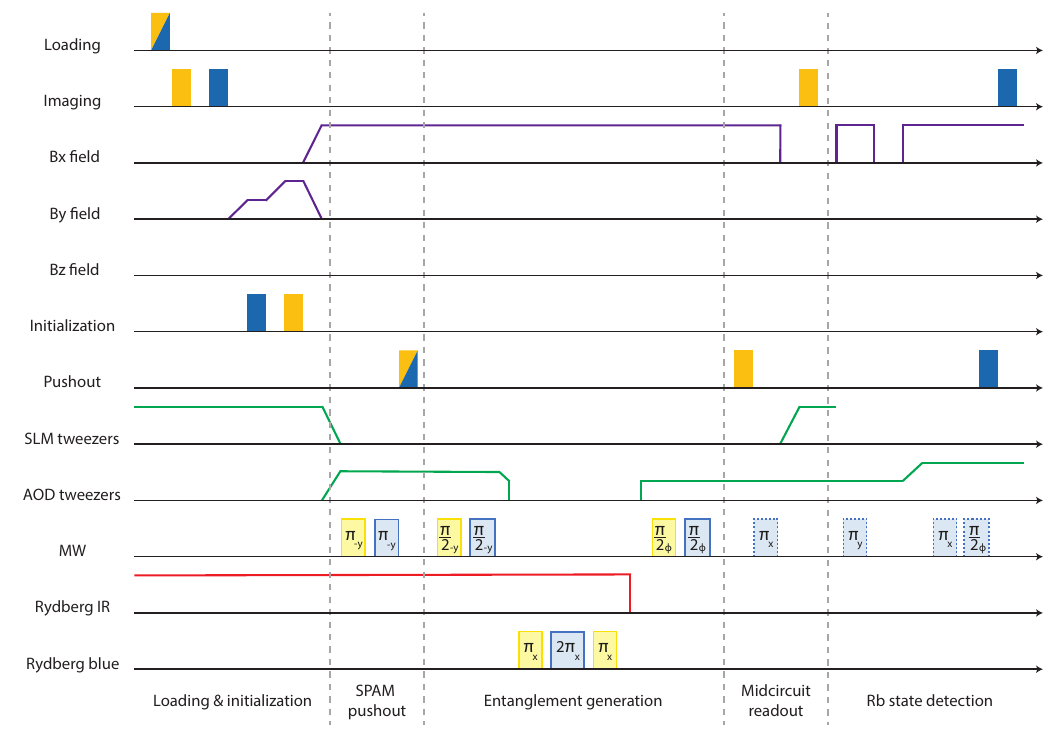}
\caption{\textbf{Experimental sequence}. In each experimental cycle, a dual-species magneto-optical trap is loaded for 30~ms from an atomic beam, followed by polarization-gradient cooling \cite{singh2022}. After loading into species-selective tweezer arrays, atoms trapped in weak out-of-plane SLM traps are removed using a pushout pulse. Atoms are then initialized into $\ket{1}_{\mathrm{Rb}}\,\textrm{=}\,\ket{F\,\textrm{=}\,2,m_{F}\,\textrm{=}\,0}$ and $\ket{1}_{\mathrm{Cs}}\,\textrm{=}\,\ket{F\,\textrm{=}\,4, m_{F}\,\textrm{=}\,0}$ via $\pi$-polarized optical pumping. The magnetic fields are adiabatically ramped from the pumping quantization field $B_y$ to the Rydberg quantization field $B_x$ to preserve qubit population, and atoms are transferred into common-mode AOD tweezers. MW-pulses then flip the qubits from $\ket{1}$ to $\ket{0}$ to ``hide" the $\ket{1}_{\mathrm{Rb}}$ and $\ket{1}_{\mathrm{Cs}}$ populations, and a resonant pulse pushes out atoms initialized in the incorrect $m_F$ states. This converts state preparation errors to raw loss. For generating entanglement, the tweezers are switched off, and the `$\pi$-2$\pi$-$\pi$' protocol~\cite{jaksch2000fast} is applied by pulsing the blue lasers of the Rydberg transitions for both species. Finally, atoms are recaptured, and the qubit state is mapped to atom presence in the array. If required, MCR of Cs qubits is performed while dynamically decoupling Rb (dashed MW pulses). This includes quenching of the B-field and application of a pinning field from the Cs SLM tweezers during fluorescence.}
\label{fig:sequence}
\end{figure*}

The dual-species experiments are performed at a reduced trap depth of $\sim$190~µK for Rb and $\sim$460~µK for Cs. Tweezers are quenched off for up to 3~µs during which the Rydberg pulses are fired. Atom temperatures during the drop are measured to be $\sim$18~µK for Rb and $\sim$30~µK for Cs via comparison with Monte Carlo simulations. If encoded in the gr-manifold, the ramping up of tweezers pushes out atoms in the Rydberg state. Otherwise, when encoded in the hf-manifold, the recapture is followed by a resonant state-selective pushout pulse. A subsequent fluorescence image reveals the final state configuration. Since the atoms load stochastically with $\sim$55\% loading efficiency in the array, we perform the single and paired atom experiments in parallel by post-selecting the data on individual and pair loading events. Extended Data Figure~\ref{fig:sequence} provides an overview of the experimental sequence. 

The trapping light for the two species is generated by independent M Squared SolsTiS systems, and shared with the Rydberg laser system to generate the blue wavelengths required for the respective two-photon Rydberg transitions (see Sec.~\ref{subsec:RydLasers}). After the degradation of the 840.6~nm laser, a Sirah Matisse C was introduced for trapping Rb. Additionally, due to fluctuations in the laboratory conditions which affected loading and loss statistics, the sequences for Figs.~\ref{fig:groundrydberg},~\ref{fig:hyperfine} were optimized on an array of 3 pairs of atoms (the left-hand side of the array shown in Fig.~\ref{fig:schematic}a). Due to laser degradation during the final measurement (Fig.~\ref{fig:hyperfine}e), the sequence was optimized on a single pair.

\newpage

\subsection{Midcircuit readout}

To facilitate quantum non-demolition measurement of the Rb atoms, we perform midcircuit readout of the Cs atoms. Readout of the hyperfine qubit state is performed by state-selective pushout followed by the collection of atomic fluorescence on an electron-multiplying charged coupled device camera (Andor IXON 888). For typical readout, atoms are fluoresced for 15~ms. For the MCR, however, this is reduced to 7~ms so that the Rb qubits can be decoupled on a faster timescale. The quantization field is quenched from $\sim$6.65~G to $\sim$0.16~G, and the Cs SLM tweezers are ramped up to $\sim$0.9~mK to provide a pinning field. The magnetic field is quenched high and low again, and interwoven into an XYX sequence so that any coherent deleterious effects on the Rb qubits can be decoupled away (ED Fig.~\ref{fig:sequence}). To account for the time required for coil ring-down, the interpulse delay is set to $\sim$20~ms. An example fluorescence histogram is provided in Fig.~\ref{fig:hyperfine}c. We measure the discrimination fidelity \cite{singh2023} to be 0.990(3).

\subsection{Rydberg lasers}
\label{subsec:RydLasers}

\begin{figure*}[h]
\centering
\includegraphics{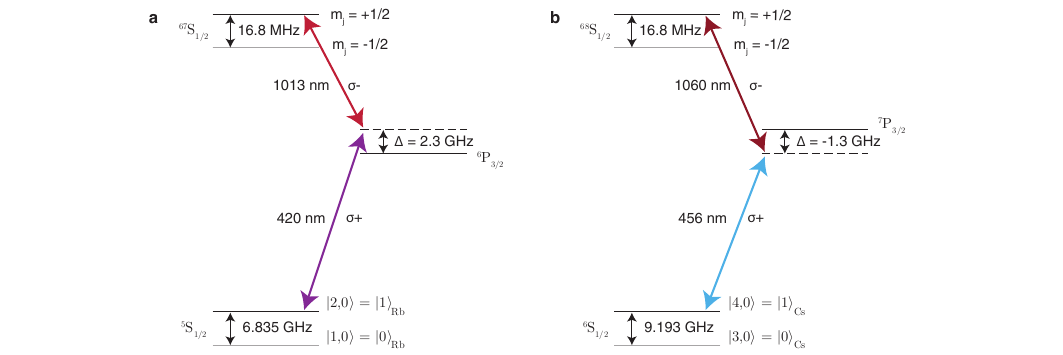}
\caption{\textbf{Two-photon Rydberg excitation}. Coherent Rydberg excitation of \textbf{a}, Rb and \textbf{b}, Cs is performed through a two-photon process using a blue and an infrared laser. Choice of intermediate state detuning is limited by available optical cavity lockpoints (spacing 3~GHz, after frequency doubling); the values of 2.3~GHz and -1.3~GHz are large enough to mitigate off-resonant scattering but small enough to achieve reasonable two-photon Rabi frequencies of 2.38~MHz and 1.86~MHz, respectively. The Rydberg states are split into two levels with an applied magnetic field of $\sim$6.65~G, enabling the selection of the $m_{j}\,\textrm{=}\,1/2$ state via laser frequency. The choice of polarization ($\sigma_{+}$, $\sigma_{-}$) further suppresses coupling to $m_{j}\,\textrm{=}\,\textrm{--}1/2$ by a factor of 3 due to Clebsch-Gordan coefficients.}
\label{fig:rydberglasers}
\end{figure*}

Rydberg states are accessed using a two-photon excitation scheme, as depicted in ED Fig.~\ref{fig:rydberglasers}. For each species, the blue light is generated by a frequency-doubling process using a Second Harmonic Generation (SHG) cavity. For the Cs $6S_{1/2}\xleftrightarrow{} 7P_{3/2}$ transition, 911.3~nm light is split from the laser used for Cs optical trapping, and converted to 455.7~nm with an M Squared ECD-X SHG. For Rb $5S_{1/2}\xleftrightarrow{} 6P_{3/2}$, a SolsTiS pumped by an M Squared Equinox generates 840.6~nm light which is doubled to 420.3~nm using an Agile Optics SHG. For the infrared transitions, a SolsTiS pumped by a Spectra-Physics Millennia eV produces 1013.0~nm light for Rb $6P_{3/2}\xleftrightarrow{} 68S_{1/2}$, and a Precilasers YFL-SF-1059 fiber seed along with a fiber amplifier produces 1060.2~nm light for Cs $7P_{3/2}\xleftrightarrow{} 67S_{1/2}$. 

\begin{figure*}[h]
\centering
\includegraphics[width=\textwidth]{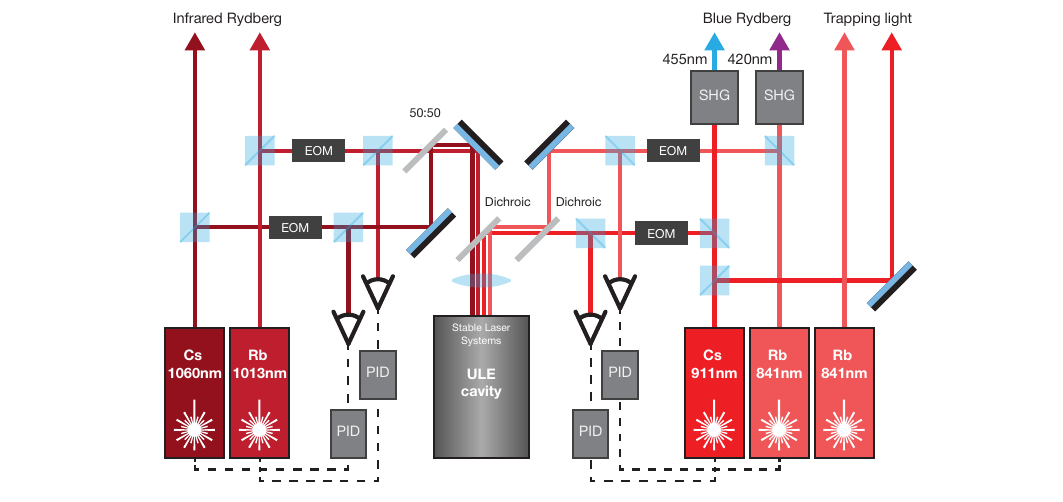}
\caption{\textbf{PDH locking setup}. Infrared Rydberg lasers (1060~nm and 1013~nm) and the fundamentals of blue lasers (911~nm and 841~nm) are simultaneously locked to a single ultra-low expansion cavity. Using the Pound-Drever-Hall technique, each laser is referenced to a high-stability cavity resonance, enabling narrow linewidths and phase coherence between all four Rydberg lasers. Both infrared lasers use an offset locking technique, and continuous tuning of the two-photon frequency is achieved by scanning the offset. Beam displacements are exaggerated in the diagram, and waveplates are omitted for clarity.}
\label{fig:laserlock}
\end{figure*}

Phase stability of the Rydberg lasers is necessary to perform coherent Rydberg operations on the atoms, and is achieved using the Pound-Drever-Hall (PDH) technique \cite{drever1983laser}. An ultra-low expansion (ULE) cavity from Stable Laser Systems (finesse $>$10,000, free spectral range 1.5~GHz) serves as a high-stability frequency reference for all four lasers, as shown in ED Fig.~\ref{fig:laserlock}. Because the cavity is anti-reflection coated for infrared wavelengths, the pre-SHG fundamentals are used for locking instead of the blue light. These two lasers use a standard PDH setup: $\sim$20~MHz phase modulation is applied using a resonant Electro-Optic Modulator (EOM) (Qubig PM7-NIR), and the reflected light from the cavity is measured on a fast photodiode. This signal is demodulated to create an error signal, which is sent through a PID controller (Vescent D2-125) and fed back to the lasers to correct frequency deviations. Since the ULE cavity only has resonances at fixed frequencies, valid lockpoints are spaced by the free spectral range, 1.5~GHz. After frequency-doubling, the blue lasers can be set in 3~GHz increments, leading to the choice of intermediate state detunings of $\Delta_{\mathrm{Rb}} \,\textrm{=}\,$2.34~GHz and $\Delta_{\mathrm{Cs}} \,\textrm{=}\,$-1.27~GHz. The two infrared lasers additionally use an `offset' PDH locking technique \cite{thorpe2008laser}. Instead of using a resonant EOM and modulating the lasers at one frequency, a broadband EOM (EOSPACE PM-0S5-05-PFA-PFA-106) is used and two modulation frequencies are applied: one at 20~MHz to create the error signal, and one adjustable from 20-730~MHz to create the offset. Scanning the offset enables frequency scanning of the laser for spectroscopic measurements (Fig.~\ref{fig:spectroscopy}).

\begin{figure*}[h]
\centering
\includegraphics[width=\textwidth]{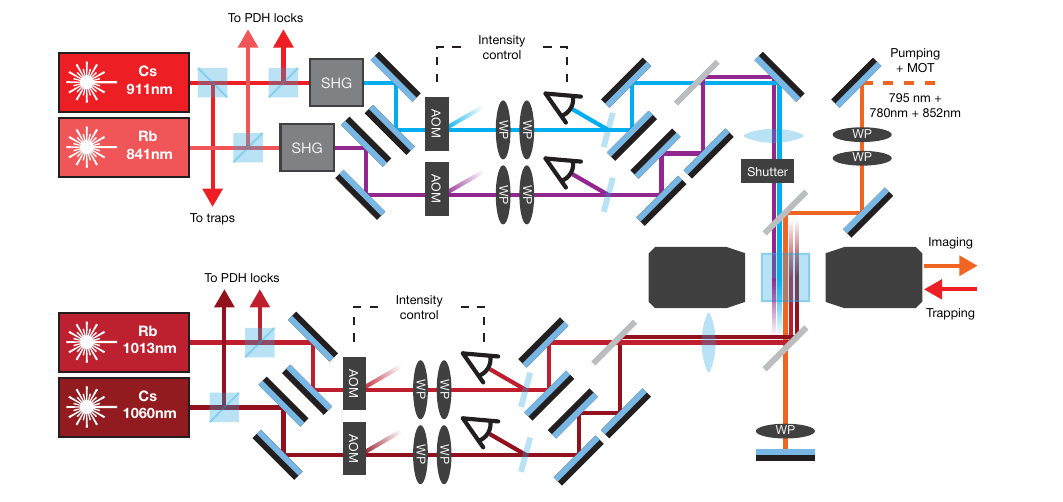}
\caption{\textbf{Rydberg laser optical layout}. Simplified schematic for the delivery of all four Rydberg lasers to the atoms. Each laser uses an AOM in a noise-eater configuration to suppress intensity noise to $<$1\%; these AOMs are also used for creating optical pulses during experimental sequences. Dichroics combine beam paths for delivery along the quantization axis, and waveplates (WP) are used to control polarization for the ($\sigma_{+}$, $\sigma_{-}$) excitation. Beams are steered between the wires of the Faraday cage, and alignment is verified on an array of atoms.}
\label{fig:laserlayout}
\end{figure*}

Extended Data Figure~\ref{fig:laserlayout} depicts how the Rydberg lasers are delivered to the atoms. Each Rydberg beam has an acousto-optic modulator (AOM) in a noise-eater configuration to reduce intensity fluctuations on the atoms to below 1\%. During experimental sequences, a sample-and-hold signal is sent to the servos controlling the AOMs, enabling pulses faster than the response time of the servo loop. The Rydberg lasers are focused into the vacuum cell, resulting in waist sizes of $\sim$30-100~µm on the atoms, based on observed variations in Rabi frequencies across different rows in a large array. Additionally, we apply UV light (365~nm, Thorlabs M365LP1) to the glass cell to stabilize the electric field environment \cite{mamat2023mitigating}. 

The four Rydberg beams are independently aligned onto the atoms by maximizing the differential Stark shift induced on qubits encoded in the hf-manifold in Ramsey-style measurements. Using the measured AC Stark shifts from the blue legs, we extract the single-photon Rabi frequencies $\Omega_{420}$ and $\Omega_{456}$ taking into account the underlying hyperfine structure of the respective $6P_{3/2}$ and $7P_{3/2}$ manifolds for Rb and Cs \cite{maller2015rydberg}. We then extract the infrared single-photon Rabi frequencies, $\Omega_{1013}$ and $\Omega_{1060}$, using the measured two-photon Rabi frequencies and intermediate state detunings. A representative set of values can be found in Table \ref{tab:exp_params} (Section \ref{sec:mesim}), taken during the hyperfine entanglement measurements.

\clearpage

\subsection{Electric Field Control}
Electric field control in the experiment is provided by six electrically-isolated field plates, shown in Fig.~\ref{fig:schematic}a. The field plates are constructed from gold-coated 316L stainless steel and form a rectangular box with outer dimensions 17~mm$\times$ 27.3~mm$\times$ 53.3~mm. In-vacuum copper wires individually connect each field plate to a multi-pin instrumentation feedthrough, allowing for external application of voltages to each of the in-vacuum field plates. This enables the set of field plates to act as a Faraday cage when all field plates are grounded, and as a set of electrodes for the creation of arbitrary electric fields on arrays of atoms. The apertures in the field plates enable optical access for laser light and high NA imaging with microscope objectives. Each aperture is covered in 15~µm diameter gold wire spaced by 0.5~mm for increased electric field suppression and greater electric field homogeneity when applying electric fields.   

\subsection{Elimination of Static Electric Field}
\begin{figure*}[h]
\centering
\includegraphics[width = 0.75\textwidth]{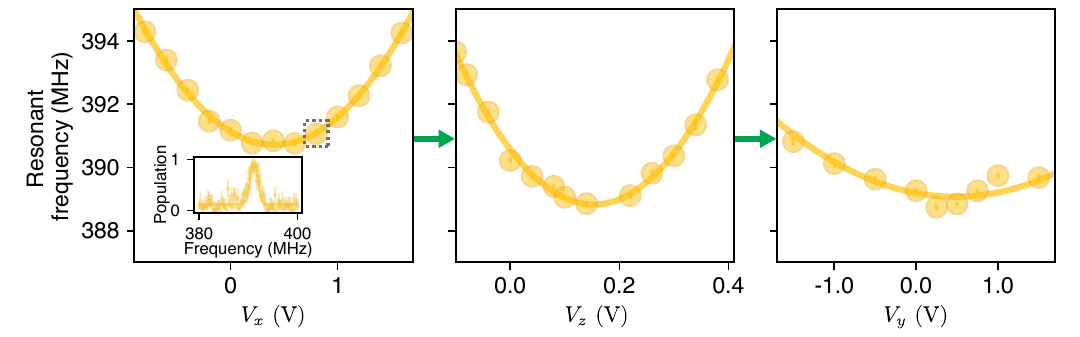}
\caption{\textbf{Cancellation of electric fields}. Environmental electric fields are nulled out using the field plates. A voltage difference $V_{x}$ is applied between the two field plates along the $x$-direction to produce a quadratic Stark shift of $\ket{67S_{1/2}}_{\text{Cs}}$ (left plot). The frequency of the 1060~nm laser is scanned with an EOM to locate the shifted Rydberg resonance. The laser is locked to the negative order of the EOM such that increasing EOM frequency corresponds to a reduction in the laser frequency. The minima of the potentials above correspond to a reduced electric field at the position of the atoms and a maximization of the energy of the Rydberg state. After locating the voltage difference $V_{x}$ that produces the minimal Stark shift on the atoms along the $x$-direction, we proceed to the $z$ (middle plot) and $y$ (right plot) directions. The inset in the first panel shows an example resonance scan of the Rydberg population as a function of the EOM frequency for the data point in the grey dashed box. These resonances are fit to Gaussian profiles to extract the center frequency.}
\label{fig:electricfieldnulling}
\end{figure*}

Stray electric fields at the position of the atoms are eliminated by applying voltage differences ($V_{x}$, $V_{y}$, and $V_{z}$) between pairs of field plates along the $x$, $y$, and $z$ directions and measuring the quadratic Stark shifts on excited Rydberg states. Extended Data Figure~\ref{fig:electricfieldnulling} illustrates our procedure using $\ket{67S_{1/2}}_{\text{Cs}}$. The resulting Stark shifts are used to determine the voltage differences that give the minima of the quadratic energy shifts, corresponding to a minimized electric field. 

The parabolic curves in ED Fig.~\ref{fig:electricfieldnulling} are fit to the following functional form: $E(v) = A(v-V_{0})^{2}+B$, where $A$, $B$, and $V_{0}$ are fitting parameters and $v$ is the applied voltage difference. The extracted fit parameters are shown in Table~\ref{table:1} for each of the three dimensions. 

\begin{table}[h!]
\centering
\begin{tabular}{c | c | c | c} 
  \textbf{Fit parameter} & $x$ & $y$ & $z$ \\ [0.5ex] 
 \hline
 $V_{0}$~(V)& $0.408 \pm 0.006$ & $0.337 \pm 0.119$ & $0.153 \pm 0.002$ \\ 
 $A$~(MHz/V$^2$)& $2.495 \pm 0.042$ & $0.533 \pm 0.109$ & $74.750 \pm 1.698$ \\
 $B$ (MHz) & $390.716 \pm 0.037$ & $389.093 \pm 0.121$ & $388.823 \pm 0.060$ \\ 
\end{tabular}
\caption{Fit parameters extracted from parabolic fits of the quadratic Stark shifts in ED Fig.~\ref{fig:electricfieldnulling}.}
\label{table:1}
\end{table}

The results from these fits can be compared to the quadratic Stark shift Hamiltonian to extract estimates for the applied electric field as a function of the experimentally applied voltage differences and to determine an estimate of the initial background electric field prior to nulling. The Stark shift Hamiltonian is given by $H = -\frac{1}{2}\alpha E^2$, where $\alpha$ is the polarizability of the $67S_{1/2}$ state of Cs and is calculated to be 465.717~MHz cm$^2$/V$^2$ using the Alkali Rydberg Calculator \cite{vsibalic2017arc}. From our comparisons, we determine that the applied electric fields as a function of the applied voltage differences $V_{x}$, $V_{y}$ and $V_{z}$ (in Volts) are approximately:
\begin{equation}
\begin{aligned}
E_{x} &= a_{x}V_{x}, \textrm{where } a_{x} = 0.1035\pm 0.0009 \textrm{ cm}^{-1}\\
E_{y} &= a_{y}V_{y}, \textrm{where } a_{y} = 0.048\pm 0.005\textrm{ cm}^{-1}\\
E_{z} &= a_{z}V_{z}, \textrm{where } a_{z} = 0.567\pm 0.006\textrm{ cm}^{-1}.
\end{aligned}
\end{equation}
From these equations, we determine that the initial electric field on the atoms prior to nulling is $\vec{E}_{\text{initial}} \,\textrm{=}\, (E_{x},E_{y},{E_{z}}) \,\textrm{=}\, (-0.0423 \pm 0.0007, -0.016 \pm 0.006, -0.087 \pm 0.001) $ (V/cm), corresponding to a total electric field of $|\vec{E}_{\text{initial}}| \,\textrm{=}\, 0.098\pm0.002$~V/cm. The electric field is largest along the $z$-direction, which corresponds to the pair of field plates with the largest apertures and consequently the lowest suppression of external electric fields. From the measured rms voltage noise of 3.5~mV on our voltage control lines, we further determine that we are able to zero out electric fields at the position of the atoms to better than 6~mV/cm.

\subsection{Effect of Static Electric Fields on F{\"o}rster Interaction}

\begin{figure*}[h]
\centering
\includegraphics[width = 0.7\textwidth]{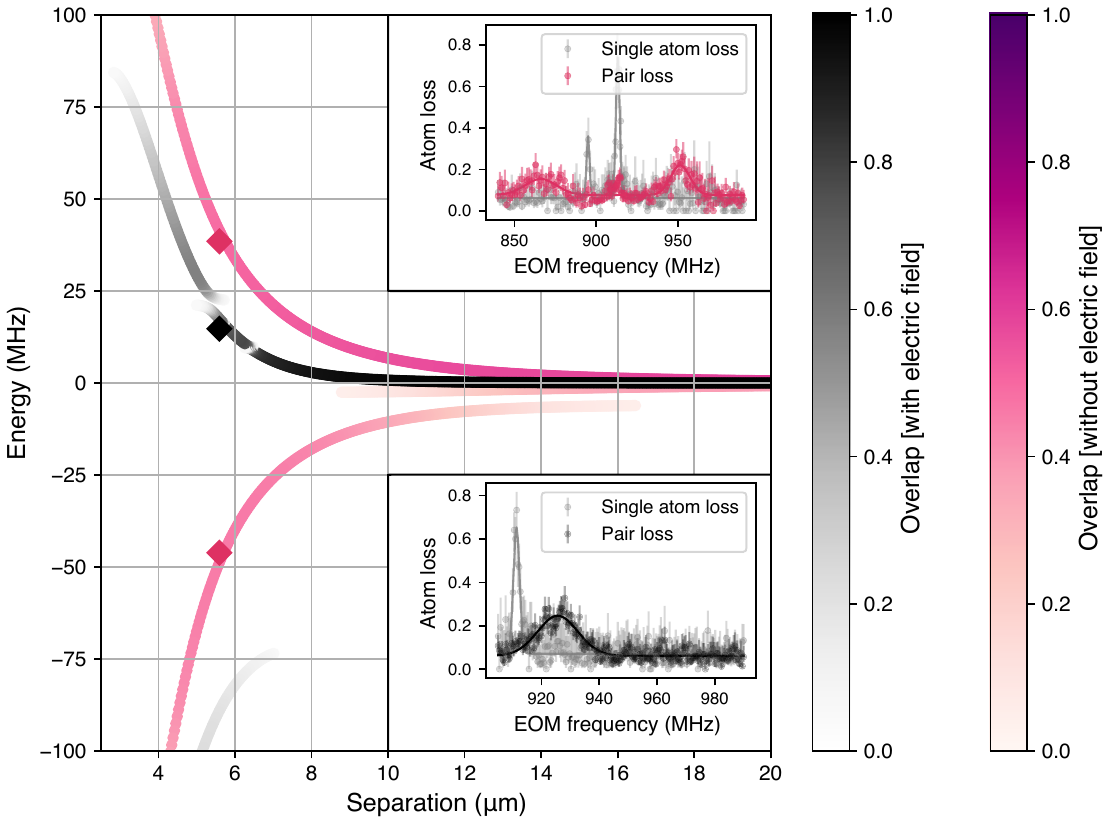}
\caption{\textbf{Effect of $\vec{E}_{\text{initial}}$ on F{\"o}rster resonance}. 
Interactions resulting from the F{\"o}rster resonance are strongly affected by background electric fields, necessitating cancellation. The pink curves show the dominant eigenstates of the near-resonant F{\"o}rster interaction arising from excitation to $\ket{68S_{1/2}}_{\mathrm{Rb}}\ket{67S_{1/2}}_{\mathrm{Cs}}$ with zero electric field and a bias magnetic field of $\sim$6.65~G along the interatomic axis. The black curves show the same system, but with an electric field corresponding to the initially measured electric field $\vec{E}_{\text{initial}}$ prior to cancellation. The colormaps show the overlap of the energy eigenstates with the non-interacting pair state. The black diamond data point indicates the measured interaction strength of 14.3(3)~MHz at a distance of 5.6~µm prior to nulling of the electric field. The associated bottom inset shows the measured pair-excitation resonance (black) for this point and the non-interacting Rb Rydberg resonance (grey) as a function of the EOM frequency on the 1013~nm laser. After nulling the electric field, we measure two pair-excitation resonances shown by the red diamond data points. The associated top inset shows the measured pair-excitation resonances (pink) for these points and the non-interacting Rb Rydberg resonances (grey). }
\label{fig:electricfieldinteractions}
\end{figure*}

Static electric fields strongly affect interspecies F{\"o}rster interactions. To illustrate this, we plot in ED Fig.~\ref{fig:electricfieldinteractions} the energy eigenvalues arising from excitation to $\ket{68S_{1/2}}_{\mathrm{Rb}}\ket{67S_{1/2}}_{\mathrm{Cs}}$ as a function of the separation between the atoms (using the Pairinteraction software \cite{Weber2017}). We plot two versions of the interaction: one with zero electric field (in pink) and one with the measured initial electric field $\vec{E}_{\text{initial}}$ prior to nulling (black). The colormaps indicate the overlap of the eigenstates with the non-interacting pair state. With the presence of the $\sim$0.1~V/cm electric field, the interaction between the two dominant eigenstates of the near-resonant F{\"o}rster pair collapse into a weaker, shorter-length interaction shown by the black curve.

\subsection{Extraction of $C_6$ and $C_3$ Coefficients}
Theoretical values of the $C_6$ coefficients for the Rb-Rb, Cs-Cs, and Rb-Cs van der Waals interactions and the $C_3$ coefficient for the Rb-Cs F{\"o}rster interaction are calculated using the Pairinteraction software \cite{Weber2017} at a bias magnetic field of 6.65~G along the interatomic axis. The $C_{6}$ coefficients are determined by fitting both the theoretical and experimental values for the van der Waals interaction strength $V(R)$ to the function $V(R) \,\textrm{=}\, C_{6}/R^{6}$ to make a direct comparison. The experimental values of the intraspecies van der Waals interactions are measured by applying a 5~µs Rydberg laser pulse on the atoms and fitting the resulting two-atom loss events as a function of laser frequency to the sum of two Voigt profiles (corresponding to the two-photon bare resonance and the four-photon resonance to the doubly excited state). For this four-photon resonance process we extract the interaction strength via the relation $\Delta E = V/2$ \cite{bernien2017probing}.

The theoretical value of the $C_{3}$ coefficient for the Rb-Cs F{\"o}rster pair is determined by taking the difference in energy between the two eigenstates with largest overlaps with the initial non-interacting pair state and fitting the result to $V(R) \,\textrm{=}\, \delta(1+C_{3}/(\delta^{2} R^{3}))$, where $\delta$ is the F{\"o}rster defect. The analogous experimental value of the $C_{3}$ coefficient is determined by taking the difference in energy between the measured pair of resonances and fitting the result to the same functional form. 

Alongside the $C_{3}$ and $C_{6}$ values quoted in the main text, we extract the F{\"o}rster defect for $\ket{68,67}_{\mathrm{Rb, Cs}}$ to be 10(1)~MHz, in agreement with the value of 9.0(2)~MHz predicted by Pairinteraction for our bias field of $\sim$6.65~G \cite{Weber2017}.

\subsection{Choice of Rydberg states for F\"orster physics}

\subsubsection{Rydberg interaction landscape}
Here we motivate the choice of Rydberg states used in our experiments. We focus on states which exhibit asymmetry between inter:intra-species interactions, which we will capture by an asymmetry parameter:
\begin{equation}
    \zeta =  \frac{\text{V}_{\text{Rb-Cs}}}{\sqrt{\text{V}_{\text{Rb-Rb}}\text{V}_{\text{Cs-Cs}}}},
    \label{eq:asymmetry}
\end{equation}
where $\text{V}_{i-j}$ denotes the interaction strength between a pair of atoms of species $i$, $j$ for some choice of Rydberg states, interatomic distance ($R$),  electric and magnetic field, and orientation of the interatomic axis with respect to the quantization axis (which we set as $\theta\,\textrm{=}\,0$ throughout this discussion). A large asymmetry parameter could be leveraged for mediated multi-qubit gate protocols and many-body physics \cite{muller2009mesoscopic}. 

As discussed in the main text, asymmetry arises due to the existence of F\"orster resonances \cite{browaeys2016experimental}. Rydberg-Rydberg interactions occur through atomic transition dipole moments which couple opposite parity states. At interatomic spacings $>$3-4~µm, the pair-state eigenvectors of the interaction dipole-dipole Hamiltonian are typically far detuned from the bare (non-interacting) pair-state, which results in a second-order van der Waals interaction: the bare pair-state is a good eigenstate of the interaction Hamiltonian, but has a shifted eigenenergy \cite{browaeys2016experimental}. However, when there coincidentally exists another nearly degenerate pair-state, a resonant dipole-dipole interaction re-emerges. 

For pairs of atoms of the same species excited to identical $S$-states --- a class of states accessible by our global two-photon excitation scheme --- F\"orster resonances are not predicted. However, such resonances \textit{are} predicted when considering interspecies combinations with differing principle quantum numbers \cite{beterov2015rydberg}. Near a resonance, the relative $1/R^{3}$ scaling of the interspecies dipole-dipole interaction compared to the $1/R^{6}$ of the intraspecies vdW interaction causes asymmetry to increase with increasing $R$. 

To identify Rydberg states which maximize asymmetry properties, we use the open-source Pairinteraction Python software package \cite{Weber2017} to calculate the interaction strengths $\text{V}_{i-j}$ for various combinations of Rydberg states of the form $\ket{n_\text{Rb}S_{1/2}, n_\text{Cs}S_{1/2}} \,\textrm{:=}\, \ket{n_\text{Rb},n_\text{Cs}}$, where $n$ denotes the principal quantum number. The magnetic quantum numbers of the chosen pair state influence the resonant interaction strength up to an angular factor \cite{beterov2015rydberg}. Here, we work with the total magnetic quantum number $m\,\textrm{=}\,1$. We focus on states around the value $n\,\textrm{=}\,70$, as this is the range accessible with our 1060.2~nm fiber laser. This range was chosen as it enables a sizeable $\zeta$ while maintaining a reasonable blockade radius for ease of array generation. We note that, given the relative scaling of each interaction type with principle quantum number (dipole-dipole: $\propto n^{4}$, vdW: $\propto n^{11}$) asymmetry generally increases at lower $n$. While omitted from the analysis here, we have thus identified the states $\ket{46,48}_\text{{Rb, Cs}}$ and $\ket{48,51}_\text{{Rb, Cs}}$ of interest for future work. These states would enable significant asymmetry ($\zeta >$ 15) at distances close to 4~µm. 

\begin{figure*}[h]
\centering
\includegraphics[width = \textwidth]{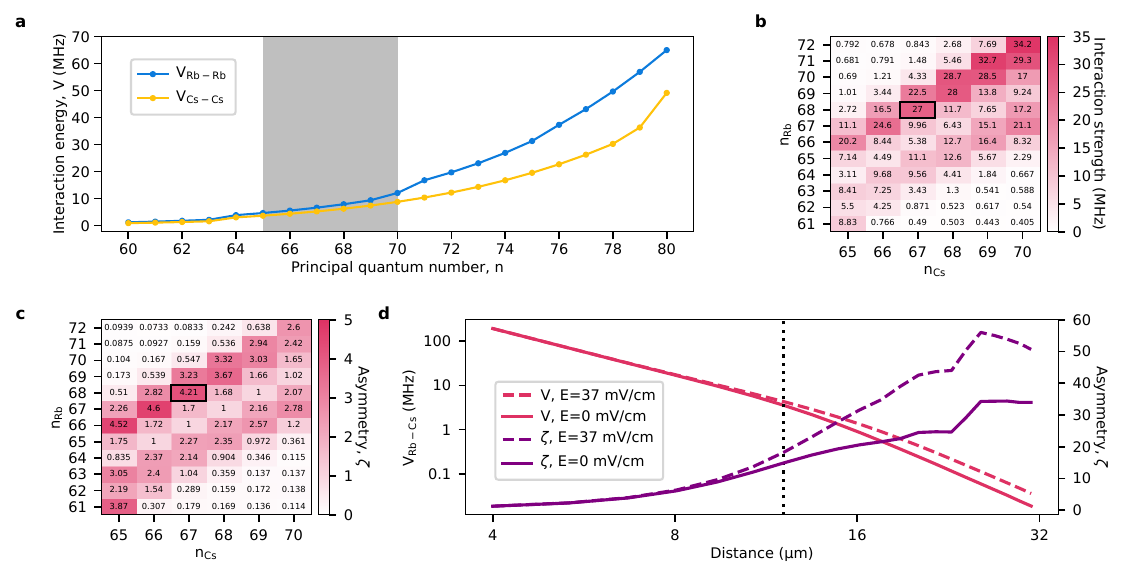}
\caption{\textbf{Rydberg state landscape for quantum information processing}. \textbf{a}, Theoretically calculated intraspecies interaction strengths at 6.65~G external magnetic field. The van der Waals interaction at an interatomic spacing of 7~µm is plotted for various principal quantum numbers of Rb (blue) and Cs (yellow), where the interaction is between atoms excited to states $\ket{nS_{1/2}, m_{j}\,\textrm{=}\,1/2}$. The shaded region shows the range of experimentally accessible Cs states. \textbf{b}, The interspecies Rb-Cs interaction strength at $B\,\textrm{=}\,$6.65~G for the principal quantum numbers of Cs shown in the shaded region in panel a. The state highlighted in the black box is $\ket{68,67}_\text{{Rb, Cs}}$ which is used for the experiments in the main text. \textbf{c}, The asymmetry parameter at zero field as calculated using Eq.~\ref{eq:asymmetry}. The highlighted box shows the state with maximal asymmetry chosen for our experiments. \textbf{d}, Stark tuning of the Rb-Cs interaction strength. The asymmetry parameter at electric fields of 0 and 37~mV/cm is plotted on the twin axis. The dotted vertical line at 12.1~µm indicates the crossover radius beyond which the interaction transitions from a $1/R^{3}$ scaling to a $1/R^{6}$ potential.}
\label{fig:rydbergstates}
\end{figure*}

\subsubsection{Numerical studies}

For the range of $n$ levels studied, we evaluate the interactions at a spacing of 7~µm. Here we expect interspecies interactions on the order of 10~MHz, which gives strong Rydberg blockade for ground-Rydberg Rabi frequencies of $\sim$1-2~MHz where good coherent control can be achieved. We perform our calculations considering the 6.65(3)~G bias field and assume that the background electric field is nulled using the Faraday cage electrodes. In general, F\"orster resonances are sensitive to both fields, and the exact asymmetry parameter will depend on how well the choice of fields suppresses the energy defect between the near-degenerate pair states \cite{ravets2015measurement}. In our analysis, we define the interaction strength between two Rydberg atoms in an initial pair state $\ket{\psi}$ as:
\begin{equation}
    V_{i-j} = \sum_{\phi} |\bra{\phi}\ket{\psi}|^{2} |U_{\phi}|
\end{equation}
where $U_{\phi}$ is the energy shift of the new eigenstates $\phi$ of the system and the summation is taken over all such eigenstates. This is a reasonable proxy for the probability of double Rydberg excitations in a pair of atoms, i.e. it gives an estimate of the blockade strength.

Extended Data Figure~\ref{fig:rydbergstates}a presents the scaling of $\text{V}_{\text{Rb-Rb}}$ and $\text{V}_{\text{Cs-Cs}}$ with $n$, showing the expected polynomial dependence. In ED Fig.~\ref{fig:rydbergstates}b we plot $\text{V}_{\text{Rb-Cs}}$ for the range of accessible Cs states. The principal quantum numbers of Rb are restricted to those for which there exists a nearby Rydberg-pair state with an energy defect in the 10~MHz range. Taking these values, we evaluate the asymmetry parameter as defined in Eq.~\ref{eq:asymmetry} (ED Fig.~\ref{fig:rydbergstates}c).  We can see clearly that for the state $\ket{68,67}_\text{{Rb, Cs}}$ the asymmetry parameter is near maximal within the range considered, and gives a stronger blockade strength than $\ket{67,66}_\text{{Rb, Cs}}$. 

As an outlook, we finally consider the enhancement of asymmetry at larger distances by electrical tuning of the $\ket{68S_{1/2}, 67S_{1/2}}_{\text{Rb, Cs}} \xleftrightarrow{} \ket{67P_{1/2}, 67P_{3/2}}_\text{{Rb, Cs}}$ F\"orster resonance to degeneracy. Numerically, we find that this can achieved by inverting the magnetic field such that $B_{z}\,\textrm{=}\,$-6.65~G, and then applying a small electric field of $E_{z}\,\textrm{=}\,$37~mV/cm. Extended Data Figure~\ref{fig:rydbergstates}d shows $\text{V}_{\text{Rb-Cs}}$ and $\zeta$ as a function of distance, both with and without the application of the electric field. At zero electric field, the interaction scales as $1/R^{3}$ at small spacings up to a crossover radius of $\sim$10~µm. At spacings larger than the crossover radius the interaction falls off as $1/R^{6}$, as $\text{V}_{\text{Rb-Cs}} < \delta$, the F\"orster defect. Under application of the electric field, the interaction is tuned to resonance ($\delta \rightarrow$ 0) and the inverse cubic scaling extends further. One would expect that, after tuning, the asymmetry would continue to scale polynomially with distance. However, for this choice of state, an interesting feature is that the initial pair state can dipole couple to superposition states of $\ket{m_{j}\,\textrm{=}\,1/2,m_{j}\,\textrm{=}\,1/2}_{\text{Rb, Cs}}$ and $\ket{m_{j}\,\textrm{=}\,\textrm{--}1/2,m_{j}\,\textrm{=}\,3/2}_{\text{Rb, Cs}}$. One superposition results in a strong $1/R^{3}$ interaction, but the other superposition is minimally interacting at short distances \cite{ravets2014coherent}. At larger distances, the Zeeman effect is typically dominant over the Rydberg-Rydberg interaction. This effectively reduces the blockade strength at larger distances. For future work targeting large asymmetry, a judicious choice of pair state (e.g a F\"orster resonance that couples two $\ket{j\,\textrm{=}\,1/2,j\,\textrm{=}\,1/2}_{\text{Rb, Cs}}$ states) can avoid this effect. One such case is predicted for $\ket{59,57}_\text{{Rb, Cs}}$.

\subsection{SPAM correction}

In this work, we build on methods described by McDonnell et al. \cite{mcdonnell2022demonstration} and Ma et al. \cite{ma2023high} to develop a robust method for the correction of state-preparation and measurement (SPAM) errors that occur during our Bell state generation and characterization circuit. Our approach is based on two key concepts: (1) the conversion of state preparation errors to atom loss and (2) the mapping of Bell state characterization observables onto `bright-bright' measurement outcomes. With high probability, this approach captures experimental runs in which two atoms were loaded into a given gate site, were prepared in the desired hyperfine state, survived the gate protocol (including the transfer between the `loading' and `science' arrays), and ended in a particular hyperfine basis state $\ket{j,k}$, ($j,k \in \{0, 1\}$). We then use a separate characterization circuit --- in which the single-qubit MW gates, the tweezer off-time, and the Rydberg laser pulses are omitted --- to capture the base probability of loading two atoms, preparing the desired initial state, and successfully transferring between the loading and science arrays. By normalizing the measurement outcomes by this base probability, we can evaluate the two-qubit gate fidelity in the absence of SPAM errors. 

\subsubsection{Conversion of state-preparation errors to atom loss}

After loading, atoms are optically pumped into the states $\ket{1}_\mathrm{Rb}$, $\ket{1}_\mathrm{Cs}$. Due to nonidealities in this process (e.g. residual vector light-shifts, polarization impurity, magnetic field noise), and loss processes (from the initial image, the transfer between trapping arrays, and background gas collisions), the resulting state of each atom is:
\begin{equation}
    \rho_{\mathrm{pump}} = p_\mathrm{g} \ket{1}\bra{1} + \sum_{m \neq 0} p_\mathrm{e, m} \ket{F, m}\bra{F, m} + p_\mathrm{l} \ket{l}\bra{l}.
\end{equation}
Here, $p_\mathrm{g}$ is the success (`good') probability, $p_\mathrm{e}\,\textrm{=}\,\sum_m p_{\mathrm{e,m}}$ is the `erroneous' ($F\,\textrm{=}\,\{2,4\}, m_{F} \neq 0$) population, and $p_\mathrm{l}$ captures loss ($\ket{l}$ is the `loss' state). We assume that repump light applied after optical pumping removes all population in the Rb $F\,\textrm{=}\,1$ and Cs $F\,\textrm{=}\,3$ hyperfine states. To remove the contributions of unpumped atoms ($p_\mathrm{e}$) to the circuit dynamics, we convert this population to loss (erasure). A microwave $\pi$-pulse converts $\ket{1}_\mathrm{Rb}$ and $\ket{1}_\mathrm{Cs}$ to $\ket{0}_\mathrm{Rb}$ and $\ket{0}_\mathrm{Cs}$, before the $F\,\textrm{=}\,2$ and $F\,\textrm{=}\,4$ manifolds are pushed out with high probability (Rb: 0.99(1), Cs: 0.97(1)). We refer to this process as state-preparation pushout (`SP-pushout'). Up to small $\pi$-pulse errors which would slightly reduce $p_{\mathrm{g}}$, the resulting state is ideally:
\begin{equation}\label{eqn:simple}
    \rho_{\mathrm{init}} = p_\mathrm{g} \ket{0}\bra{0} + p_\mathrm{L} \ket{l}\bra{l},
\end{equation}
with $p_{\mathrm{L}} \,\textrm{=}\, p_{\mathrm{l}} + p_{\mathrm{e}}$. 

We note that there is a small probability for atoms to survive pushout. In this case, we assume that the atoms are distributed in the $F=\{1,3\}$, $m_{F}\neq$ 0 states, and undergo minimal further dynamics as the microwave- and Rydberg-transitions are far-detuned. We instead have:
\begin{equation}
    \rho'_{\mathrm{init}} = p_\mathrm{g} \ket{0}\bra{0} + p_{\mathrm{s}}p_\mathrm{e} \ket{a}\bra{a} + p_\mathrm{L'} \ket{l}\bra{l}.
\end{equation}
where $\ket{a}$ denotes `atom present', i.e. population which will contribute to the final fluorescence measurement. In practice, with $p_{\mathrm{e}} \sim 5-10\%$ and $p_{\mathrm{s}} \sim 1-3\%$, the population of $\ket{a}$ should remain $\ll1\%$. SP-pushout is used in all measurements presented in Figs.~\ref{fig:groundrydberg} and \ref{fig:hyperfine} of the main text, but not in Figs.~\ref{fig:schematic} and \ref{fig:spectroscopy}.

\subsubsection{Mapping measurement bases to `bright, bright'}

We now turn to Bell state characterization via `bright, bright' measurements. The state-preparation circuit (Fig.~\ref{fig:hyperfine}b of the main text, up to dashed line) ideally creates the canonical Bell state $\ket{\Psi^{+}} \,\textrm{=}\, \frac{1}{\sqrt{2}}(\ket{0_{\mathrm{Cs}},0_{\mathrm{Rb}}} + \ket{1_{\mathrm{Cs}},1_{\mathrm{Rb}}})$, up to a Cs single-qubit phase, which can be compensated in later operations. Hereafter we will drop the subscript notation for the atomic species, and will maintain the ordering Cs, Rb. The fidelity of the prepared state can be evaluated via the relation $\mathcal{F}_{\mathrm{Bell}} \,\textrm{=}\, (P_{00} + P_{11} + P_{c})/2$, where $P_{00}$ and $P_{11}$ are the populations for the observables $\mathcal{O}_{00} \,\textrm{=}\, \ketbra{0,0}{0,0}$ and $\mathcal{O}_{11} \,\textrm{=}\, \ketbra{1,1}{1,1}$, and $P_{c}$ captures the Bell state coherences, given by the observable  $\mathcal{O}_\mathrm{c} \,\textrm{=}\, \ketbra{0,0}{1,1} + \ketbra{1,1}{0,0}$.

At the end of the sequence, the spin-state $\ket{0(1)}$ is converted to atom presence (loss). Thus, $P_{00}$ manifests directly as the probability $p_{\mathrm{bb}}$ (`bright, bright'), whereas mapping $P_{11}$ to $p_{\mathrm{bb}}$ is achieved by applying additional $\pi$-pulses to both qubits prior to pushout.  To obtain $P_{c}$, the observable $\mathcal{O}_{\mathrm{c}}$ must be mapped to $\mathcal{O}_{00}$, which is realized by applying $\pi/2(\phi)$ pulses to both qubits. Specifically, in this case we measure a probability $p^{\mathrm{c}}_{\mathrm{bb}} \,\textrm{=}\, \textrm{Tr}(U \rho U^{\dagger} \mathcal{O}_{00})$, with $U \,\textrm{=}\, \mathrm{exp}[-i\frac{\pi}{4} \sigma_{\phi, \phi}]$, where $\sigma_{\phi, \phi} \,\textrm{=}\, [\cos{(\phi)} \ \sigma^{\mathrm{Cs}}_{x} + \sin{(\phi)} \ \sigma^{\mathrm{Cs}}_{y}] + [\cos{(\phi)} \ \sigma^{\mathrm{Rb}}_{x} + \sin{(\phi)} \ \sigma^{\mathrm{Rb}}_{y}]$.

Akin to Ref. \cite{ma2023high}, this can be related to the measurement of the observable:
\begin{equation}
    \mathcal{O}_{\phi} = U^{\dagger}\mathcal{O}_{00}U = \frac{1}{4}\begin{bmatrix} 1 & -ie^{-i\phi} & -ie^{-i\phi} & -e^{-2i\phi}\\ ie^{i\phi} & 1 & 1 & -ie^{-i\phi}\\ ie^{i\phi} & 1 & 1 & -ie^{-i\phi} \\ -e^{2i\phi} & ie^{i\phi} & ie^{i\phi} & 1 \end{bmatrix}.
\end{equation}

The coherences of the Bell state manifest in $p^{\mathrm{c}}_{\mathrm{bb}}$ as the term oscillating with frequency $2\phi$, and the value $P_{\mathrm{c}}$ is given by 4A, where A is the amplitude of that oscillating term (Fig.~\ref{fig:hyperfine}b of the main text).

\subsubsection{Readout imperfections}

Ideally, the raw Bell state fidelity would be found via $\widetilde{\mathcal{F}}^{\mathrm{raw}}_\mathrm{Bell} \,\textrm{=}\, (\widetilde{P}_{00} + \widetilde{P}_{11} + \widetilde{P}_{c})/2$, where $\widetilde{P}_{ij}$ are the measured populations, giving a SPAM-corrected fidelity of $\widetilde{\mathcal{F}}_\mathrm{Bell} \,\textrm{=}\, \widetilde{\mathcal{F}}^{\mathrm{raw}}_\mathrm{Bell}/p^{\mathrm{SPAM}}_{\mathrm{bb}}$, with $p^{\mathrm{SPAM}}_{\mathrm{bb}}$ the probability to measure $\ket{\mathrm{bright,bright}}$ after the gate-free SPAM characterization circuit. However, this picture can be complicated by read-out errors.

These fall under: (1) `discrimination' errors; an atom is assigned as present  (absent) when there is none (one), or (2) `mapping' errors; either (a) an atom was left in $\ket{a}$ after attempted erasure ($P_{a}$), (b) an atom in the state $\ket{1}$ survives the  final pushout pulse ($p_{\mathrm{fp}}$), or (c) an atom is lost during the gate sequence or imaging. The discrimination fidelity of our standard (15~ms) imaging histograms is $>$0.999, and thus those errors have minimal impact. For mapping errors, atom loss will result in a reduction of $p_{\mathrm{bb}}$ (assigned as a gate error), but `false-positive' events may lead to an overestimate. 

To understand these effects, we modify the derivation of Ref. \cite{ma2023high} to account for independent error rates for the two species, and for the residual population in $\ket{a}$, $P^{\mathrm{Rb/Cs}}_{a}$. This population will appear in measurements of both $\widetilde{P}_{00}$ and $\widetilde{P}_{11}$, but not $\widetilde{P}_{c}$, as $P_{a}$ does not result in an oscillating signal. For small $P_{a}$ and $p_{\mathrm{fp}}$, keeping only first-order terms in them (and products thereof), we find the relationship between the underlying ($P_{ij}$) and measured ($\widetilde{P}_{ij}$) populations:

\begin{equation}\label{eq:popn}
P_{00} + P_{11} \approx \frac{(\widetilde{P}_{00} + \widetilde{P}_{11}) - (p^{\mathrm{Cs}}_{\mathrm{fp}}p^{\mathrm{Rb}}_{\mathrm{tp}}+p^{\mathrm{Cs}}_{\mathrm{tp}}p^{\mathrm{Rb}}_{\mathrm{fp}})}{(p^{\mathrm{Cs}}_{\mathrm{tp}}p^{\mathrm{Rb}}_{\mathrm{tp}}-p^{\mathrm{Cs}}_{\mathrm{tp}}p^{\mathrm{Rb}}_{\mathrm{fp}}-p^{\mathrm{Cs}}_{\mathrm{fp}}p^{\mathrm{Rb}}_{\mathrm{tp}}+p^{\mathrm{Cs}}_{\mathrm{fp}}p^{\mathrm{Rb}}_{\mathrm{fp}})} - 2(P^{\mathrm{Cs}}_{a} + P^{\mathrm{Rb}}_{a}), 
\end{equation}
\begin{gather}
    P_{\mathrm{c}} = \frac{\widetilde{P}_{\mathrm{c}}}{p^{\mathrm{Cs}}_{\mathrm{tp}}p^{\mathrm{Rb}}_{\mathrm{tp}}+p^{\mathrm{Cs}}_{\mathrm{fp}}p^{\mathrm{Rb}}_{\mathrm{fp}}-p^{\mathrm{Cs}}_{\mathrm{tp}}p^{\mathrm{Rb}}_{\mathrm{fp}}-p^{\mathrm{Cs}}_{\mathrm{fp}}p^{\mathrm{Rb}}_{\mathrm{tp}}},
\end{gather}

where $p_{\mathrm{tp}}, p_{\mathrm{fp}}$ refer to true- and false-positive probabilities (namely atoms in $\ket{1}$ which survived pushout). Note that the second term in Eq.~\ref{eq:popn} is an upper bound for the quantity ($P_{0a} + P_{1a} + P_{a0} + P_{a1} + P_{aa}$), i.e. ($P_{00} + P_{11}$) is lower bounded.

The raw Bell state fidelity is lower-bounded by:
\begin{equation}
\mathcal{F}^{\mathrm{raw}}_\mathrm{Bell} = (P_{00}+P_{11}+P_{\mathrm{c}})/2 =  \frac{(\widetilde{P}_{00}+\widetilde{P}_{11}+\widetilde{P}_{\mathrm{c}}) - (p^{\mathrm{Cs}}_{\mathrm{tp}}p^{\mathrm{Rb}}_{\mathrm{fp}}+p^{\mathrm{Cs}}_{\mathrm{fp}}p^{\mathrm{Rb}}_{\mathrm{tp}})}{2(p^{\mathrm{Cs}}_{\mathrm{tp}}p^{\mathrm{Rb}}_{\mathrm{tp}}+p^{\mathrm{Cs}}_{\mathrm{fp}}p^{\mathrm{Rb}}_{\mathrm{fp}}-p^{\mathrm{Cs}}_{\mathrm{tp}}p^{\mathrm{Rb}}_{\mathrm{fp}}-p^{\mathrm{Cs}}_{\mathrm{fp}}p^{\mathrm{Rb}}_{\mathrm{tp}})} - (P^{\mathrm{Cs}}_{a}+P^{\mathrm{Rb}}_{a}).
\end{equation}

A further lower bound is found with both $p_{\mathrm{tp}} \,\textrm{=}\, 1$. For our measured $\widetilde{\mathcal{F}}^{\mathrm{raw}}_\mathrm{Bell}\,\textrm{=}\,0.49(2)$, and our estimated values of $p^{\mathrm{Cs}}_{\mathrm{fp}}\,\textrm{=}\, 0.03$, $p^{\mathrm{Rb}}_{\mathrm{fp}}\,\textrm{=}\, 0.01$, we predict a reduction of $\mathcal{F}^{\mathrm{raw}}_\mathrm{Bell}$ by 4$\cdot$10$^{-4}$. While the population in $\ket{a}$ is not directly measurable in our experiment, a product of the raw pumping infidelity and the pushout error gives a simple estimate of $P^{\mathrm{Cs}}_{a}\,\textrm{=}\,0.003$, $P^{\mathrm{Rb}}_{a} \,\textrm{=}\,0.0005$. The extracted, SPAM-corrected Bell state fidelity of $\widetilde{\mathcal{F}}_{\mathrm{Bell}}$ \,\textrm{=}\, 0.69(3) is in good agreement with the prediction of our numerical model ($\mathcal{F_{\mathrm{Bell}}}$ \,\textrm{=}\, 0.73), giving further confidence to the treatment made here. In future work targeting high-fidelity operations, rigorous gate fidelity estimates can be achieved using interleaved randomized benchmarking \cite{magesan2012efficient} or gate-set tomography \cite{nielsen2021gate}.

\subsubsection{Mitigation of SP errors for blockade measurement}

For the blockade measurement presented in Fig.~\ref{fig:groundrydberg}b of the main text, state preparation errors on the Rb qubit will result in an absence of blockade. To mitigate this effect, we employ SP-pushout to convert SP errors to loss. After applying the variable time Cs drive, we apply a second $\pi$-pulse to the Rb qubit to de-excite it, and post-select the data on retention of the Rb atom. For the remainder of the measurements presented in Fig.~\ref{fig:groundrydberg} of the main text, post-selection is not performed.

\subsubsection{SPAM correction of eye diagrams}
For the `eye diagram' presented in Fig.~\ref{fig:hyperfine}a of the main text, SP-pushout is employed on both atomic species. The Cs qubit state is mapped to the bright state prior to measurement, such that the data can be post-selected on retention of the Cs atom. This largely suppresses the effects of SPAM errors on that qubit, allowing the reliable (post-selected) preparation of a chosen input state on the Cs atom. We then correct for SPAM errors on the Rb qubit using the fitted contrast of $A\,\textrm{=}\,0.872(14)$ obtained when applying the same sequence in the absence of the Rydberg light or the tweezer drop.

For the eye diagram presented in Fig.~\ref{fig:hyperfine}d, we apply the same correction process on the Rb qubit (again using $A\,\textrm{=}\,0.872(14)$), but the Cs qubit is prepared in a superposition state, resulting in a random measurement outcome. To suppress the effects of SP errors on the Cs qubit, we associate the state $\ket{0}$ with `dark', as both $\ket{0}$ and atom loss correspond to scenarios in which blockade should not occur. The reduced contrast of Fig.~\ref{fig:hyperfine}d compared with Fig.~\ref{fig:hyperfine}a then results from two main factors. 

First, the Rb qubit undergoes dephasing during the $\sim$60~ms time associated with MCR. This effect could be improved by reducing the MCR time (for example, imaging at high field using a single cycling imaging beam \cite{martinez2017fast,kwon2017parallel} and using qCMOS camera technologies \cite{bluvstein2023logical}). Second, the imperfect Cs pushout (0.97(1)) before the MCR results in projection errors on the Rb qubit, mapping it onto the opposite ($\pi$-out-of-phase) curve, which reduces the amplitude when averaging over the outcomes. 

\subsubsection{SPAM correction of QND measurement}
For the QND measurements (Fig.~\ref{fig:hyperfine}e of the main text), the Rb qubit state is mapped to the bright state prior to measurement, such that the data can be post-selected on retention of the Rb atom (reliably preparing a chosen input state).  
The best estimate for the QND readout fidelity is then given by correcting for SP errors on the Cs (auxiliary) qubit. We again employ SP-pushout, but, to remain agnostic to the input state on the data qubit, we do not change the Cs measurement based on the expected outcome (i.e. we do not add any additional $\pi$-pulses on Cs to enable post-selection on $\ket{\mathrm{bright, bright}}$ outcomes). 

Immediately after taking the QND-measurement data, the Cs loss probability without application of the gate was measured to be $P^{\mathrm{Cs}}_{\mathrm{SP}}\,\textrm{=}\,0.22(2)$. Correcting the data for the loss-probability associated with state preparation, we find: 
$\mathcal{F}^{\mathrm{SP}}\,\textrm{=}\,(1-P_\mathrm{loss})/(1-P_{\mathrm{SP}})$. We thus find $P(\mathrm{0}|\mathrm{0})\,\textrm{=}\,0.78(2)$, $P(\mathrm{1}|\mathrm{1})\,\textrm{=}\,0.73(3)$, as presented in the main-text.

We note that post-selection on retention of the Rb qubit suppresses error mechanisms of the two-qubit gate associated with loss of the Rb atom (e.g. loss due to the drop or an imperfect Rydberg $2\pi$-pulse), which might lead to an over-estimate in the fidelity of the auxiliary-based read-out. This effect only plays a role when the Rb qubit is in the $\ket{1}$ state. Our numerical model predicts a SPAM-free $P(1|1)\,\textrm{=}\,0.788$ when post-selecting, or $P(1|1)\,\textrm{=}\,0.766$ without post-selection ($P(0|0)\,\textrm{=}\,0.775$ for both cases). The error-budget for this sequence is dominated by the Cs $T_{2, \mathrm{gr}}^{*}$; without this effect, we predict SPAM- and post-selection-free $P(1|1)\,\textrm{=}\,0.955$ and $P(0|0)\,\textrm{=}\,0.970$.

\subsection{QND-ness of auxiliary-based measurement}

Alongside the ability of the auxiliary qubit to extract information from the data qubit, an important feature of an auxiliary-based measurement is that it preserves the data qubit state. To assess the `QND-ness' for the Rb atom, we evaluate the fraction of the data for which it was retained (i.e. survived SP-pushout, application of the gate, and the final measurement pushout) and compare this to the bare Rb atom loss in the absence of the QND-measurement sequence, which was found to be $P^{\mathrm{Rb}}_{\mathrm{SP}}\,\textrm{=}\,0.11(1)$ immediately after the taking of the data. That is, we evaluate $P^{\mathrm{QND}}\,\textrm{=}\,P_{\mathrm{ret}}/(1-P_{\mathrm{SP}})$. We find an average QND-ness $P^{\mathrm{QND}}_{0/1}\,\textrm{=}\,0.94(2)$ ($P^{\mathrm{QND}}_{0}\,\textrm{=}\,0.97[2]$,  $P^{\mathrm{QND}}_{1}\,\textrm{=}\,0.91[2]$). The QND-ness for the state $\ket{0}$ is likely limited by background atom loss during the sequence time, alongside losses during the tweezer-off time. For the state $\ket{1}$, the additional errors are likely due to a combination of the additional pair of microwave pulses and the fact that the Rb atom undergoes Rydberg excitation in this case. 

\subsection{Master equation simulations}\label{sec:mesim}

To analyze ground-Rydberg qubit dynamics and the entangling operations implemented on hyperfine qubits, we perform modeling using a Master equation solver based on the QuTiP Python toolkit \cite{johansson2013qutip}. For each atomic species, our model incorporates the following levels, \{$\ket{0}$, $\ket{1}$, $\ket{i}$, $\ket{r}$, $\ket{r'}$, $\ket{l}$\}. Here $\ket{0}$ and $\ket{1}$ are hyperfine clock-states, $\ket{i}$ is the intermediate state for the two-photon excitation process, $\ket{r}$ and $\ket{r'}$ are the desired ($m_{j}\,\textrm{=}\,1/2$) and erroneous ($m_{j}\,\textrm{=}\,\textrm{--}1/2$) Rydberg states, and $\ket{l}$ is an effective `leakage' state encompassing atom loss and decay to states outside the qubit subspace.

A summary of the relevant decoherence mechanisms and their modeling is given in the following sections. The timescale and functional form for each process is either derived from an independent characterization measurement, or taken from literature values. Each mechanism is incorporated into the model via the relevant jump operator. Throughout the evolution, the amplitude of these operators is varied depending on the system Hamiltonian at a given point in time (for example, if the qubit is idle or being driven). 

The relevant experimental parameters are given in Table \ref{tab:exp_params}, while the decoherence timescales and associated functional forms are given in Table \ref{tab:decoherence}. The values for the Rabi frequencies and dephasing timescales correspond to a representative dataset taken during the hyperfine entanglement measurements.

\begin{table}[!h]
    \centering
    \begin{tabular}{c|c|c}
        \textbf{Parameter} & \textbf{Value Rb} & \textbf{Value Cs}  \\ [0.5ex]
        \hline
        Rydberg $n$ & 68 & 67\\
        Rydberg $m_{j}$ splitting [MHz] & \multicolumn{2}{c}{18.6(1)} \\
        Magnetic field, B$_{z}$ [G] & \multicolumn{2}{c}{6.65(3)} \\
        Hyperfine splitting [GHz] & 9.1926 \cite{SteckRb} & 6.8347 \cite{SteckCs} \\
        Hyperfine Rabi frequency [kHz] & 5.62(2) & 8.55(4) \\
        Intermediate-state detuning* [GHz] & -2.34 & 1.27 \\
        Diff. Stark shift induced by g-e drive [MHz] & 6.42(1) & 3.05(1) \\
        Single-photon Rabi frequency, g-e [MHz] & 199.6(2) & 131.4(2) \\
        Single-photon Rabi frequency, e-r [MHz] & 55.80(3) & 35.97(5) \\
        Two-photon Rabi frequency, g-r [MHz] & 2.380(5) & 1.860(7) \\
        Rydberg interaction strength, $V_{\mathrm{eff}}$** [MHz] & \multicolumn{2}{c}{24} \\
        Temperature at drop [µK] & 18 & 30
    \end{tabular}
    \caption{Relevant experimental parameters, as used in numerical simulations. *: Defined with respect to the ground-intermediate state transition frequency. **: Effective strength, see text.}
    \label{tab:exp_params}
\end{table}

\begin{table}[!h]
    \centering
    \begin{tabular}{c|c|c|c}
        \textbf{Parameter} & \textbf{Value Rb} & \textbf{Value Cs} & \textbf{Decay shape}  \\ [0.5ex]
        \hline
        Rydberg state lifetime $T_{1,r}$ [µs] & 138 \cite{vsibalic2017arc} & 126 \cite{vsibalic2017arc} & Exponential\\
        Intermediate state $T_{1,e}$ [ns] & 129 \cite{vsibalic2017arc} & 154 \cite{vsibalic2017arc} & Exponential \\
        Atom loss (2.5~µs drop) & 0.004(6) & 0.009(7) & Exponential \\
        $T_{2, \mathrm{gr}}^{*}$ (idle) [µs] & 0.81(5) & 0.9(1) & Exponential \\
        $T_{2,\mathrm{gr}}^{\mathrm{d}}$ (driven) [µs] & 4.1(4) & 4.6(6) & Gaussian \\
        $T_{2, \mathrm{hf}}^{*}$ (idle) [ms] & 6.7(5) & 0.26(3) & Gaussian\\
        $T_{2,\mathrm{hf}}$ (single-echo) [ms] & 52(7) & 12(1) & Gaussian\\
        $T_{2,\mathrm{hf}}^{\mathrm{Ryd}}$ (Rydberg driving) [µs] & 1.62(9) & 2.5(2) & Gaussian
    \end{tabular}
    \caption{Decoherence parameters used in the error models. Values with error-bars are extracted from fits to independent measurements using the given decay function.}
    \label{tab:decoherence}
\end{table}

\subsection*{Error sources}

\subsubsection{Blockade strength}

In this work, we choose to utilize Rydberg states exhibiting an interspecies F\"orster resonance. Here, the concept of blockade strength as usually considered for the van der Waals regime is more complex. In particular, for the F\"orster resonance, there are  now multiple pair-state eigenstates of the interaction Hamiltonian which have non-negligible overlap with the bare pair-state $\ket{68,67}_\mathrm{Rb,Cs}$. The double excitation probability is given by a sum over the transition amplitudes from singly-excited states to these pair-states. For simplicity, we capture this in the Master equation via an effective blockade strength, introduced as a density-density operator on the bare pair-state: $V\ketbra{r,r}$. We further assume uniform V for all combinations of $\ket{r}$, $\ket{r'}$.

To calculate the effective blockade strength, $V$, we take an approach inspired by de Léséleuc, Weber et al. \cite{de2018accurate}. Under a simplified Hamiltonian comprising a single ground state and direct ground-Rydberg coupling, we solve the Schr\"odinger equation across a driving period of 10~µs, taking into account the 10 Rydberg pair-states with the largest overlap to the bare state, which we identify using the Pairinteraction software package \cite{Weber2017}. Considering a single-atom Rabi frequency of 2~MHz, we find a time-averaged $p_{rr}\,\textrm{=}\,0.002$ (the sum of the populations in doubly-excited states). We find an equivalent time-averaged $p_{rr}$ occurs for a single excitation channel with an effective interaction strength of $V\,\textrm{=}\,$24~MHz.

\subsubsection{Atomic state lifetimes}

Within the two-photon excitation scheme, there is a probability to populate the intermediate state, which decays on a fast timescale (129~ns [Rb], 154~ns [Cs] \cite{vsibalic2017arc}). We make a worst-case assumption here that all scattering results in a leakage error. Similarly, we account for finite Rydberg state lifetimes of 138~µs and 126~µs for Rb and Cs respectively, which we calculate using ARC \cite{vsibalic2017arc}. This is broken down into radiative-decay and black-body channels of 342~µs, 231~µs (Rb) and 300~µs, 217~µs (Cs). For simplicity, here we again assume that all decay results in leakage errors. We separately account for scattering due to the IR excitation beams which remain on for the duration of the $\pi$-$2\pi$-$\pi$ scheme. Our modelling predicts that the presence of this light results in an effective $T^{\mathrm{Rb}}_{1,\mathrm{R}}\,\textrm{=}\,$113~µs, $T^{\mathrm{Cs}}_{1,\mathrm{R}}\,\textrm{=}\,$92~µs.

\subsubsection{Atom loss}

In the analysis of the hyperfine-basis two-qubit gate, we neglect state-preparation and measurement errors (see `SPAM correction' section for description of the correction of the measured data). Thus, in this case, atom loss only enters from the finite drop time of the tweezers for Rydberg excitation. We implement this as a decay of all levels to the $\ket{l}$ state during the Rydberg pulses.

For the ground-Rydberg experiments, we do not SPAM correct, but do perform SP-pushout prior to the measurements in order to avoid unwanted excitation dynamics from incorrectly prepared atoms. We model this by an initial population $\epsilon_\mathrm{SP}$ in the $\ket{l}$ state, which is treated as a dark state in the final measurement. 

\subsubsection{Rydberg detection}

For the ground-Rydberg experiments, we account for a finite Rydberg detection error, $\epsilon_\mathrm{Ryd}$, which is captured by the measurement operators:
\begin{gather}
\Pi_\mathrm{bright} = \Pi_{0} + \Pi_{1} + \epsilon_\mathrm{Ryd}(\Pi_{r} + \Pi_{r'}), \\
\Pi_\mathrm{dark} = (1-\epsilon_\mathrm{Ryd})(\Pi_{r} + \Pi_{r'}) + \Pi_{i} +  \Pi_{l}.
\end{gather}
Here, $\Pi_\mathrm{bright}, \Pi_\mathrm{dark}$ are the projection operators for the `bright' and `dark' measurement outcomes.

To find $\epsilon_\mathrm{Ryd}$, we then assume that the height of the Rabi oscillation is only limited by finite coherence and Rydberg state detection. Noting that, after SP-pushout, the atom loss probabilities take values in the range $\{\epsilon_\mathrm{SP},1\}$, we can correct the data as $P^\mathrm{SP}_\mathrm{retained} \,\textrm{=}\, (1-P_\mathrm{loss})/(1-\epsilon_\mathrm{SP})$. We then extrapolate the peak of the Rabi oscillations (minimal $P^\mathrm{SP}_\mathrm{retained}$) back to $t=0$ using the fitted decay function, which is our best estimate for $\epsilon_\mathrm{Ryd}$ \cite{levine2019parallel}. Based on the fitted decay of our ground-Rydberg Rabi oscillations, we estimate that the $\pi$-pulse fidelity is $>$99\% on both species.

\subsubsection{Dephasing mechanisms}

Our two-qubit gate fidelity is limited predominantly by dephasing mechanisms. Here, we outline the relevant timescales and decay forms as used for the Master equation jump operators. We also elucidate our understanding of the mechanisms behind them.

The $\pi$-$2\pi$-$\pi$  gate scheme is particularly sensitive to idle dephasing of the control qubit (Cs $T_{2, \mathrm{gr}}^{*}$). We measure exponential-type dephasing with a characteristic timescale of 0.9(1)~µs via a ground-Rydberg Ramsey measurement. The measured atom temperature of $\sim$30~µK (Doppler shift of $\omega_\mathrm{D} \sim$2$\pi\cdot$54~kHz) corresponds to an expected dephasing timescale of ${T_{\mathrm{D}} \,\textrm{=}\, \sqrt{2} \omega_{\mathrm{D}}^{-1}}$ $\sim$4~µs. We thus hypothesize that the dephasing time is primarily limited by phase noise of the Rydberg lasers, which will be characterized in detail in future work. 

Beyond $T_{2, \mathrm{gr}}^{*}$, there is dephasing which occurs during the two-photon drive on each species. This manifests in two forms. First, on the ground-Rydberg qubit, we measure a damping of the Rabi oscillations ($\Omega^{\mathrm{Rb}}_{\mathrm{gr}} \,\textrm{=}\, 2\pi\cdot$2.380(5)~MHz, $\Omega^{\mathrm{Cs}}_{\mathrm{gr}} \,\textrm{=}\, 2\pi\cdot$1.860(7)~MHz) with characteristic timescales of $T_{2,\mathrm{gr}}^{\mathrm{d,Rb}}\,\textrm{=}\,$4.1(4)~µs and $T_{2,\mathrm{gr}}^{\mathrm{d,Cs}}\,\textrm{=}\,$4.6(6)~µs ($N^{\mathrm{gr}}_{\mathrm{cycles}} \,\textrm{=}\, 9.8[9], 8.6[1.1]$). Second, within a Ramsey-type measurement with variable blue pulse length, we observe dephasing of the hyperfine qubit due to the differential AC Stark-shift between the $\ket{0}$ and $\ket{1}$ states. The measured Stark shifts and dephasing timescales are $\delta^{\mathrm{Rb}}_{\mathrm{AC,blue}}\,\textrm{=}\,$6.42(1)~MHz, $T_{2,\mathrm{hf}}^{\mathrm{d,Rb}}\,\textrm{=}\,$1.62(9)~µs and $\delta^{\mathrm{Cs}}_{\mathrm{AC,blue}}\,\textrm{=}\,$3.05(1)~MHz, $T_{2,\mathrm{hf}}^{\mathrm{d,Cs}}\,\textrm{=}\,$2.5(2)~µs ($N^{\mathrm{hf}}_{\mathrm{cycles}} \,\textrm{=}\, 10.4[6], 7.6[6]$). For both the gr and hyperfine dephasing, we find Gaussian decay ($\mathrm{exp}[-(t/T)^{2}]$). 

For the Stark-shift measurements, the blue light is far off-resonant in the absence of the IR drive, and so phase noise is not expected to play a role. Assuming quasi-static, normally-distributed intensity noise with standard deviation $\sigma_{\mathrm{I, blue}}$, the expected number of coherent Stark-shift oscillations ($1/e$ decay) can be found to be: $N^{\mathrm{hf}}_{\mathrm{cycles}} \,\textrm{=}\, [\sqrt{2}\pi\sigma_{I}]^{-1}$. Inverting the measured $N^{\mathrm{hf}}_{\mathrm{cycles}}$, we obtain estimates for $\sigma^{\mathrm{Rb}}_\mathrm{I,blue}\,\textrm{=}\,0.022(1)$ and $\sigma^{\mathrm{Cs}}_\mathrm{I,blue}\,\textrm{=}\,0.030(2)$. This is inconsistent with independent measurements of the intensity fluctuations of the Rydberg pulses as measured on a fast photodiode, for which we find $\sigma_{\mathrm{I}}$ to be less than 1\% for all Rydberg beams. We thus hypothesize that the this decay comes from movement of the atoms within the laser intensity profiles. This could be alleviated by beam shaping \cite{ebadi2021quantum}. Balancing the single-photon Rabi frequencies (presently limited by IR laser power) would lead to a $\sim$4x reduction in the differential Stark shifts, or a $\sim$4x improvement in $N^{\mathrm{hf}}_{\mathrm{cycles}}$. This would likely also improve the ground-Rydberg coherence, as it can be shown that --- in the case of equal intensity fluctuations on the legs of the two-photon drive --- two-photon detuning errors arising from intensity noise are minimized for balanced single-photon Rabi frequencies. Finally, the implementation of fast hyperfine control via Raman driving would enable echoing of the differential Stark shift \cite{levine2019parallel}.

For the damping of the Rydberg Rabi oscillations, intensity noise from both legs plays a role, as does laser phase noise. It is challenging to independently extract a value for the infrared intensity noise via a differential Stark-shift measurement, as the Stark shifts are comparable to the inverse hyperfine coherence time (sub-kHz). Neglecting phase noise allows estimation of $\sigma^{\mathrm{Rb}}_\mathrm{I,IR}\,\textrm{=}\,0.040(5)$ and $\sigma^{\mathrm{Cs}}_\mathrm{I,IR}\,\textrm{=}\,0.043(8)$ via Monte Carlo modeling. However, these values are also inconsistent with photodiode measurements and may well be an overestimate. Detailed studies of the phase noise spectrum of the four Rydberg lasers will be required to further understand the limits to both the driven and undriven ground-Rydberg coherence \cite{de2018analysis}.

\subsubsection{Error budget for the two-qubit gate}

Combining the mechanisms discussed above, we give an error budget for the two-qubit gate in Table \ref{tab:error_budget}. 

\begin{table}[!h]
    \centering
    \begin{tabular}{c|c|c}
        \textbf{Parameter} & \textbf{Independent infidelity} & \textbf{Cumulative BSF} \\ [0.5ex]
        \hline
        Intermediate-state scattering & 0.0128 & 0.9872 \\
        Idle gr dephasing & 0.2032 & 0.7861 \\
        Differential Stark dephasing  & 0.0435 & 0.7541 \\
        Atom loss & 0.0120 & 0.7451 \\
        Driven gr dephasing & 0.0085 & 0.7395 \\
        Rydberg decay & 0.0031 & 0.7372 \\ 
        Finite blockade & 0.0018 & 0.7359 \\
        Coupling to $\ket{r'}$ & 0.0008 & 0.7327  \\
    \end{tabular}
    \caption{Bell state generation error budget. `Independent infidelity' is the simulated Bell state infidelity in the presence of only that error channel. To obtain these infidelities in the absence of intermediate-state scattering we separately use a 5-level model (removing $\ket{i}$) with a direct $\ket{1}\xleftrightarrow{}\ket{r/r'}$ coupling at the two-photon Rabi frequency). In this case we neglect $\ket{0}\xleftrightarrow{}\ket{r/r'}$ coupling. `Cumulative BSF' is the Bell state fidelity including all error channels up to that entry in the table, calculated using the full 6-level model.}
    \label{tab:error_budget}
\end{table}

\subsubsection{Ground-Rydberg simulations}

\begin{figure*}[h]
\centering
\includegraphics[width=0.79\textwidth]{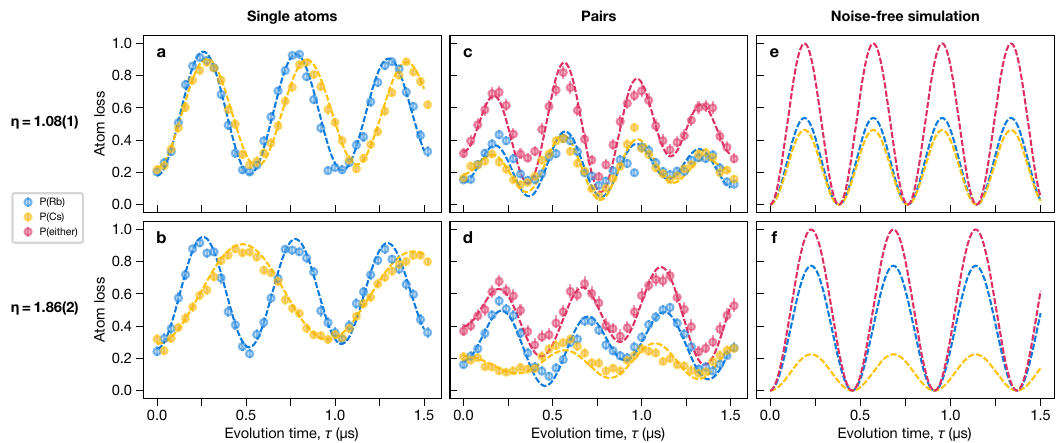}
\caption{\textbf{Interspecies Rydberg dynamics}. \textbf{a-d} Replicated from Fig.~\ref{fig:groundrydberg}. \textbf{a}, \textbf{b}, Rydberg Rabi frequencies are determined by driving the ground-Rydberg transition on single atoms. \textbf{c}, \textbf{d} When simultaneously driving Rb-Cs pairs, Rabi oscillations are observed with enhanced frequency and decreased excitation probability per atom. The dynamics are affected by SPAM errors, but these effects are captured by simulations including the same errors (dashed lines). \textbf{e}, \textbf{f}, In the absence of SPAM errors, Rb and Cs are expected to oscillate at the same frequency, but with different amplitudes, determined by the ratio $\eta\,\textrm{=}\,\Omega_{\textrm{Rb}}/\Omega_{\textrm{Cs}}$ (methods).}
\label{fig:grSimulation}
\end{figure*}

The values used in the ground-Rydberg numerical simulations (Fig.~\ref{fig:groundrydberg}c-e) are given in Table \ref{tab:gr_sims}. In ED Fig.~\ref{fig:grSimulation} we reproduce the simultaneous driving results presented in Fig.~\ref{fig:groundrydberg} of the main text alongside noise-free simulations which show the ideal dynamics in the absence of SPAM or decoherence errors.

\begin{table}[!h]
    \centering
    \begin{tabular}{c|c|c|c|c}
        & \multicolumn{2}{c|}{$\eta$ = 1.08(1)} & \multicolumn{2}{c}{$\eta$ = 1.86(2)} \\ \hline \textbf{Parameter} & \textbf{Rb} & \textbf{Cs} & \textbf{Rb} & \textbf{Cs} \\
        \hline
        $\epsilon_\mathrm{SP}$ & 0.18(2) & 0.23(1) & 0.26(1) & 0.29(1) \\
        $\epsilon_\mathrm{Ryd}$ & 0.07(3) & 0.15(3) & 0.07(3) & 0.15(3) \\
        $\Omega$ [MHz] & 1.92(1) & 1.78(1) & 1.93(1) & 1.04(1) \\
        $T_{2,\mathrm{gr}}^{\mathrm{d}}$ (driven) [µs] & 3.6(9) & 5.1(1.7) & 3.6(9) & 3.4(4)  \\
    \end{tabular}
    \caption{Parameters used for the ground-Rydberg qubit simulations (Fig.~\ref{fig:groundrydberg}c-e of the main text). $\epsilon_\mathrm{SP}$ is found as the offset, $o$, when fitting the single-atom loss probability during (unblockaded) Rabi oscillations to the function: $f(t) \,\textrm{=}\, o + A (1 - \cos(\Omega(t-t_0)) \exp[-(t/T_{2,\mathrm{gr}}^{\mathrm{d}})^{2}])$.}
    \label{tab:gr_sims}
\end{table}

\newpage
\subsection{Collective driving of intraspecies pairs}

\begin{figure*}[h]
\centering
\includegraphics{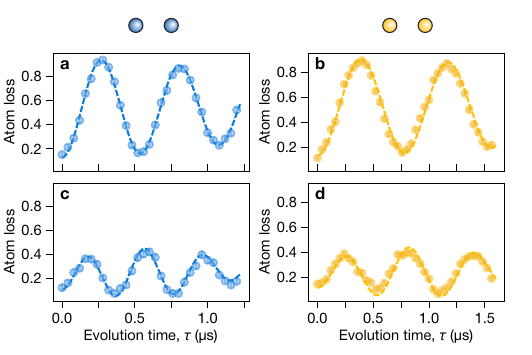}
\vspace{-10pt}
\caption{\textbf{Intraspecies Rydberg Rabi oscillations} Similar to the simultaneous driving of Rb-Cs pairs in Fig.~\ref{fig:groundrydberg}c-e, pairs of Rb-Rb or Cs-Cs atoms can be driven in the blockade regime. \textbf{a}, \textbf{b}, Per-site atom loss when loading single atoms, showing Rabi oscillations between the ground and Rydberg states. \textbf{c}, \textbf{d}, When driving pairs of atoms, we recover the standard result: the Rabi frequency is enhanced by a factor of $\sqrt{2}$ compared to the single-atom case, and the per-atom excitation probability is no larger than 1/2. These results show that a strong blockade is still achieved for intra- as well as interspecies operations, provided the atoms are sufficiently close to each other (here 5.6~µm).}
\label{fig:intraspecies}
\end{figure*}

Alongside the interspecies simultaneous driving results presented in Fig.~\ref{fig:groundrydberg} of the main text, we performed analogous experiments on Rb-Rb and Cs-Cs pairs. For these cases, the Rabi frequency is equal for both atoms. Comparing single-atom Rabi oscillations (ED Fig.~\ref{fig:intraspecies}a,b) and collective oscillations of pairs of atoms (ED Fig.~\ref{fig:intraspecies}c,d) we observe both the expected $\sqrt{2}$ enhancement of the collective Rabi frequency and the 2x suppression of the per-site excitation probability.
\vspace{-10pt}

\subsection{Simultaneous driving with independent Rabi frequencies}

We consider the Hamiltonian:
\begin{equation}
    \mathcal{H} =  \frac{\hbar}{2}\Omega_{1} (\ket{gg}\bra{rg} + \ket{gr}\bra{rr} + h.c.)
    + \frac{\hbar}{2}\Omega_{2} (\ket{gg}\bra{gr} + \ket{rg}\bra{rr} + h.c.) + V \ketbra{rr},
\end{equation}
where $\Omega_{i}$ are the Rabi frequencies acting on qubit $i \in \{1,2\}$, and $V$ is the Rydberg interaction. In the limit $V \gg \Omega_{1,2}$ (strong blockade), the state $\ket{rr}$ is frozen out, and the Hamiltonian simplifies to:
\begin{equation}
\mathcal{H_{\mathrm{blockade}}} = \frac{\hbar}{2}\Omega_{1} (\ket{gg}\bra{rg} + h.c.) + \frac{\hbar}{2}\Omega_{2} (\ket{gg}\bra{gr} + h.c.).
\end{equation}
Defining a generalized $W$-like state: $|\widetilde{W}\rangle \,\textrm{=}\, \frac{\Omega^{*}_{1} \ket{rg} + \Omega^{*}_{2} \ket{gr}}{\sqrt{\Omega^{2}_{1} + \Omega^{2}_{2}}}$, and collective Rabi frequency $\widetilde{\Omega}\,\textrm{=}\,\sqrt{\Omega^{2}_{1}+\Omega^{2}_{2}}$, this becomes:
\begin{equation}
\mathcal{H_{\mathrm{blockade}}} = \widetilde{\Omega} (\ket{gg}\langle\widetilde{W}|+h.c.).
\end{equation}
That is, the system oscillates between $\ket{gg}$ and $|\widetilde{W}\rangle$ with a Rabi frequency $\widetilde{\Omega}$. For the special case of $\Omega_{1}\,\textrm{=}\,\Omega_{2}$, we retrieve the standard result of $\widetilde{\Omega} \,\textrm{=}\, \sqrt{2}\Omega_{1}$, $|\widetilde{W}\rangle\,\textrm{=}\,(\ket{rg}+\ket{gr})/\sqrt{2} \,\textrm{:=}\, \ket{W}$.

\subsection{Quantum State Transfer}
Species-selective Rydberg excitation enable simple pulse schemes for performing quantum state transfer and controlling the spatial flow of quantum information \cite{cesa2023universal}. Beginning first with a pair of Rb and Cs atoms in the $\ket{1}_{\mathrm{Rb}}\ket{1}_\mathrm{Cs}$ state, we can use a global drive on the Rb atoms to create the following superposition in the $(\ket{1}\xleftrightarrow{}\ket{r})$ qubit space: $(\alpha\ket{1}_{\mathrm{Rb}} + \beta\ket{r}_{\mathrm{Rb}})\ket{1}_{\mathrm{Cs}}$. Within the strongly-blockaded regime, a Rydberg $\pi$-pulse on the Cs atom and a subsequent $\pi$-pulse on the Rb atom can then be used to transfer the quantum information from Rb to Cs, up to a single-qubit X-rotation on the Cs qubit:
\begin{align*}
(\alpha\ket{1}_{\mathrm{Rb}} + \beta\ket{r}_{\mathrm{Rb}})\ket{1}_{\mathrm{Cs}} &= \alpha \ket{1}_{\mathrm{Rb}}\ket{1}_{\mathrm{Cs}} + \beta \ket{r}_{\mathrm{Rb}}\ket{1}_{\mathrm{Cs}}\\
\pi \text{-pulse~on~Cs}\rightarrow&= 
\alpha \ket{1}_{\mathrm{Rb}}\ket{r}_{\mathrm{Cs}} + \beta \ket{r}_{\mathrm{Rb}}\ket{1}_{\mathrm{Cs}}\\
\pi \text{-pulse~on~Rb}\rightarrow&= 
\alpha \ket{1}_{\mathrm{Rb}}\ket{r}_{\mathrm{Cs}} + \beta \ket{1}_{\mathrm{Rb}}\ket{1}_{\mathrm{Cs}}\\
&= 
\ket{1}_{\mathrm{Rb}}(\alpha \ket{r}_{\mathrm{Cs}} + \beta \ket{1}_{\mathrm{Cs}}).
\end{align*}
The experimental results of this quantum operation are shown in Fig.~\ref{fig:groundrydberg}f.

\end{appendices}

\end{document}